\documentclass[12pt]{iopart}
\usepackage{iopams}

\usepackage{setstack}

\usepackage{color}

\newlength{\bredde}
\def\slash#1{\settowidth{\bredde}{$#1$}\ifmmode\,\raisebox{.15ex}{/}
\hspace*{-\bredde} #1\else$\,\raisebox{.15ex}{/}\hspace*{-\bredde} #1$\fi}

\newcommand{\diag}{{\rm diag\,}}

\newcommand{\be}{\begin{equation}}
\newcommand{\ee}{\end{equation}}
\newcommand{\bea}{\begin{eqnarray}}
\newcommand{\eea}{\end{eqnarray}}
\newcommand{\nn}{\nonumber}
\newcommand{\sect}[1]{\setcounter{equation}{0}\section{#1}}

\def\erfc{{\mbox{erfc}}}
\def\sgn{{\mbox{sgn}}}
\def\Tr{{\mbox{Tr}}}
\def\Pf{{\mbox{Pf}}}
\def\re{{\Re\mbox{e}}}
\def\im{{\Im\mbox{m}}}

\def\La{\Lambda}

\def\al{\alpha}
\def\ka{\kappa}

\definecolor{red}{rgb}{1,0,0}
\newcommand{\eins}{\leavevmode\hbox{\small1\kern-3.8pt\normalsize1}}

\newcommand{\mcA}{\mathcal{A}}
\newcommand{\mcD}{\mathcal{D}}

\newcommand{\mcK}{\mathcal{K}}

\begin{document}
\title[Skew-OP for chiral real asymmetric random matrices]{Skew-orthogonal
Laguerre polynomials for chiral real asymmetric random matrices}

\author{G.\ Akemann$^{1,2}$, {M.\ Kieburg}$^{3}$ and M.J.\ Phillips$^1$}

\address{\it
$^1$Department of Mathematical Sciences \& BURSt Research Centre,\\
\ \ Brunel University West London, Uxbridge UB8 3PH, United Kingdom\\~\\
$^2$The Niels Bohr Institute \& The Niels Bohr International Academy,\\
\ \ Blegdamsvej 17, 2100 Copenhagen {\O}, Denmark\\~\\
$^3$Fakult\"at f\"ur
Physik, Universit\"{a}t Duisburg-Essen, 47048 Duisburg, Germany
}
\begin{abstract}
We apply the method of skew-orthogonal polynomials (SOP) in the complex
plane to asymmetric random matrices with real elements, belonging to
two different classes. Explicit integral representations valid for
arbitrary weight functions are derived for the SOP
and for their Cauchy transforms, given 
as expectation values of traces and determinants or their inverses,
respectively. 
Our proof uses the fact that the joint probability distribution function for
all combinations of 
real eigenvalues and complex conjugate eigenvalue pairs can be written
as a product. 
Examples for the SOP are given in terms of Laguerre polynomials for
the chiral ensemble (also called the non-Hermitian real
Wishart-Laguerre ensemble), both without and with the insertion of 
characteristic polynomials. Such characteristic polynomials play the
role of mass terms in 
applications to complex Dirac spectra in field theory.
In addition, for the elliptic real Ginibre ensemble we recover the SOP
of Forrester 
and Nagao in terms of Hermite polynomials.

\end{abstract}




\sect{Introduction}

Classical orthogonal polynomials (OP) are one of the principal
standard tools used to 
solve problems in Random Matrix Theory (RMT). The three classical
Wigner-Dyson ensembles 
with Gaussian elements
can be solved in terms of Hermite polynomials, whereas their chiral
counterparts require the use of Laguerre polynomials. Whilst for the symmetry
classes with unitary invariance ($\beta=2$) 
the corresponding scalar product is symmetric with the standard
weight, for the classes with
orthogonal ($\beta=1$) or symplectic ($\beta=4$) symmetry the scalar product 
becomes skew-symmetric, with the details -- including the weight -- dependent
on the symmetry class. The corresponding polynomials are then called
skew-orthogonal polynomials (SOP).
For details of all these cases we refer to
\cite{Mehta},
as well as to \cite{Ghosh} and \cite{Taro} for reviews on SOP.
What all these ensembles have in common is that each of their solutions
can be expressed using the kernel 
of the corresponding (S)OP as a building block.

The solution of RMT in terms of (S)OP is exact for finite
matrix size $N$. Moreover, when taking one of the possible large-$N$
limits in the bulk of the spectrum, or at the soft or hard edge,
the standard Plancherel-Rotach asymptotics can be used. Much
work has been done on the question of universality, i.e.\ the extent to which
the asymptotics also hold for non-Gaussian weights, see
e.g. \cite{Deift} for a review.

This setup of (S)OP has been generalised to the complex
plane in order to solve non-Hermitian RMT. However, for the standard
Ginibre ensembles the (S)OP are simply 
given in terms of monic powers (which holds for all
weights that are rotationally invariant in ${\mathbb C}$). Only when
considering so-called elliptic deformations of the Gaussian Ginibre
weight do Hermite polynomials on ${\mathbb C}$ appear, as was first
observed in \cite{PDF}. The corresponding quaternionic ($\beta=4$)
elliptic Ginibre ensemble was solved in terms of some Hermite SOP  in
\cite{Kan02}, and only very recently was the real elliptic Ginibre
case ($\beta=1$) solved in terms of another set of 
Hermite SOP \cite{Forrester07}.

The chiral counterparts of two of these ensembles were introduced in
\cite{James} 
($\beta=2$) and \cite{A05} ($\beta=4$), where they were solved in
terms of Laguerre OP on ${\mathbb C}$  
and Laguerre SOP respectively.
The obvious question then arises about the existence of Laguerre SOP for the 
chiral $\beta=1$ ensemble 
\cite{APSo,APSoII} which we answer affirmatively in this article, thereby
completing the set of classical Hermite and Laguerre (S)OP in the
complex plane for these non-Hermitian RMT.

One complication arises for the ensembles with Laguerre (S)OP in the
complex plane: due to the integration over angular variables the
elliptic ensembles that are Gaussian in terms of the matrix elements lead to 
non-Gaussian weight functions for the eigenvalues 
(for $\beta=1$ elliptic Ginibre
we have also a complementary error function). The Bessel function of the
second kind appearing here for all three $\beta=1,2,4$ 
makes the orthogonality question much more
involved. 

For this reason we first provide a new integral representation for the 
$\beta=1$ SOP, valid in both the elliptic Ginibre and chiral symmetry classes
for arbitrary weight functions. In view of earlier results for
SOP on ${\mathbb R}$ for both $\beta=1,4$ 
\cite{Eynard,GP}, as well as for complex SOP for $\beta=4$
\cite{Kan02,A05}, this representation comes very naturally. Moreover,
it was shown very recently in \cite{BHJ} that the  $\beta=1$ and $\beta=4$
Ginibre ensembles can be treated on an equal footing. We will rederive
this relation amongst these symmetry classes from a different angle. 
We will also derive a new integral representation for the
Cauchy transforms of the SOP on ${\mathbb C}$ valid for both
$\beta=1$ and $4$. 
This extends the expression for Cauchy transforms on ${\mathbb C}$
for $\beta=2$ in \cite{AP04}. 

One important ingredient necessary in order to derive these results is
the factorisation of the joint probability distribution function
(jpdf) for $\beta=1$, 
which is originally given by a sum over all possible combinations of
real and complex conjugate eigenvalues \cite{Lehmann91,Edelman97}. Such a
factorisation, which uses the symmetrisation over all eigenvalues,
might have
been expected from the fact that the partition function can be
written as a single Pfaffian over double integrals \cite{Sinclair}. 

Our integral representation allows us to derive the  $\beta=1$
Laguerre SOP on ${\mathbb C}$ in a straightforward fashion. The known 
Hermite SOP on ${\mathbb C}$ of \cite{Forrester07} also follow
easily. 
As a third and important example for our general formalism we
explicitly compute the SOP for the chiral $\beta=1$ ensemble with the
insertion of mass terms. Such insertions play a crucial role in the
application of RMT to the complex Dirac operator spectrum in Quantum
Chromodynamics (QCD)  and related field theories at small quark 
chemical potential 
in the low density phase, see e.g. \cite{A07mu} for reviews, 
as well as very recently in the
high density phase corresponding to maximal non-Hermiticity \cite{Tilo}.

For this third example we exploit the fact that averages (and ratios)
of the required 
characteristic polynomials have been computed very recently for the
non-Hermitian $\beta=1$ symmetry classes in \cite{KGII}. Together with
our interpretation of the building blocks there as the kernel, SOP and
their Cauchy transforms this completes the analogy to earlier
computations of such averages and ratios for $\beta=2$ in
\cite{AV03,AP04,Bergere} and $\beta=4$ in \cite{A05,AB}. 

This article is organised as follows.
In section \ref{MM} we recall the definition of the two RMT with real
asymmetric matrices including mass terms, their respective weight functions and
eigenvalue representations. Section \ref{jpdf} is devoted to a
factorisation proof of the jpdf of real and complex eigenvalues, where
we give two different arguments. The new integral representations for
the SOP and their Cauchy transforms are then shown in
subsections \ref{sops} and \ref{cauchy} respectively.
In section \ref{examples} we provide three explicit examples for SOP
including Hermite in subsection \ref{sHermite}, Laguerre SOP in subsection
\ref{sLaguerre}
and SOP including masses in subsection \ref{smass}. 
In {the Appendices}
we collect together short
proofs for some mathematical identities used in the text.

\sect{The Matrix Models}\label{MM}

We will show how to solve the following two matrix models of real
asymmetric matrices in terms of skew-orthogonal polynomials (SOP) in
the complex plane.

The first model is given by the chiral extension of the elliptic 
real Ginibre ensemble
\bea
\fl{\cal Z}_{ch}^{(N_f)}(\{m\})&\sim&
\int_{\mathbb{R}^{2N(N+\nu)}}
dA\,
dB
\prod_{f=1}^{N_f}\det
\left(\begin{array}{cc}
m_f\eins_{N\times N}& A\\
B^T&m_f\eins_{(N+\nu)\times (N+\nu)}\\
\end{array}
\right) \label{Zch}\\
&&\times
\exp\left[-\frac12\eta_+\Tr(AA^T+BB^T)
+\eta_-\Tr(AB^T)
\right]\ ,\nn\\
&&
\ \ \mbox{with}\ \ \eta_\pm\ \equiv\ \frac{1\pm\mu^2}{4\mu^2}
\label{mudef}
\ .
\eea
Here $A$ and $B$ are real asymmetric matrices of size
$N\times(N+\nu)$, 
and $\mu\in(0,1]$ is a non-Hermiticity parameter. The integration runs
over all independent real matrix elements of $A$ and $B$ with a flat measure.
The product of
determinants or characteristic polynomials is motivated by the
addition of $N_f$ quark flavours in applications to QCD at finite density
\cite{A07mu,Tilo}.
The model can be
written as a Gaussian two-matrix model with two independent real
asymmetric matrices $P$ and $Q$, with $A=P+\mu Q$ and $B=P-\mu Q$
(see \cite{APSoII}). In the limit $\mu\to0$ the
model reduces to the chiral Gaussian Orthogonal Ensemble (chGOE).

The second matrix model is also a generalisation of the elliptic real
Ginibre ensemble and is given by
\bea
\fl{\cal Z}^{(N_f)}_{Gin}(\{m\})&\sim&
\int_{\mathbb{R}^{N^2}}dJ
\prod_{f=1}^{N_f}|\det[J-im_f]|^2
\exp\left[-\frac{1}{1-\tau^2}\Tr(JJ^T-\tau J^2)
\right]\ .
\label{Zgin}
\eea
Here $J$ is a real asymmetric matrix of size $N^2$, and $\tau\in[0,1)$
is the non-Hermiticity parameter. We again integrate over all the
independent matrix elements of $J$. 
The model can alternatively be written as a two-matrix model with 
symmetric and anti-symmetric matrices $S$ and $A$ with
Gaussian elements, where $J=S+A\sqrt{(1-\tau)/(1+\tau)}$, see also
\cite{FKS98} for $N_f=0$.  
In the limit $\tau\to1$ the model reduces to the GOE when $N_f=0$.
The extra determinants correspond to the
imaginary mass terms coming in pairs of opposite sign. These are also 
motivated from applications to QCD with a chemical potential, 
but this time in three dimensions, see \cite{A01}.

Once we switch to an eigenvalue basis for the Dirac matrix ${\cal D}$ for
the first model, where 
\be
{\cal D}\equiv
\left(\begin{array}{cc}
\mathbf{0}_{N\times N}& A\\
B^T&\mathbf{0}_{(N+\nu)\times (N+\nu)}\\
\end{array}
\right)
\ ,
\label{Mdef}
\ee
and to the eigenvalues of the matrix $J$ for the second model, both models
can be treated along the same lines. Because the characteristic equation
for both ${\cal D}$ and $J$ is real its solutions are either real or
come in complex conjugate pairs. 
However, because of the chirality of ${\cal D}$ there is a
peculiarity here:
\be
0=\det[\Lambda \eins_{2N+\nu}-\mcD]=\La^\nu\det[\La^2 \eins_N-AB^T]=
\La^\nu\prod_{j=1}^N(\La^2-\La_j^2)\ .
\label{evrel}
\ee
Whilst the $\La_j^2$ are indeed either real or come in complex
conjugate pairs, the Dirac eigenvalues $\La_j$ are consequently
real ($\La_j^2>0$), purely imaginary ($\La_j^2<0$), or come in
quadruplets ($\pm\La_j,\pm\La_j^*$); there are also $\nu$ 
{generic zero-eigenvalues}.
For simplicity and to keep the presentation of the
two models parallel we will mainly consider changed variables
$z_j \equiv \La_j^2$ in the following.

Following \cite{Sommers2007,APSo} 
the partition functions in eqs.\ (\ref{Zch})
and (\ref{Zgin}) above can be written as follows, where the
normalisation is to be determined later, see eq. (\ref{Zprod}),
\be
\fl{\cal Z}_{2N+\chi}=
N!
\sum\limits_{n=0}^{N}\
\prod_{k=1}^{2n+\chi} \int\limits_{\mathbb{R}} dx_k
\prod_{m=1}^{N-n} \int\limits_{\mathbb{C}} d^{\,2}z_m \,
P_{2n+\chi,N-n}(x,z,z^*)\ .
\label{Zsum}
\ee
Here we sum over all the possible ways of splitting the total number
($2N+\chi$) of  
eigenvalues into $K\equiv 2n+\chi$ real eigenvalues {$\{x_k\}$} and
$M\equiv N-n$ 
complex conjugate eigenvalue pairs {$\{z_m,z_m^*\}$}. A product with
an upper limit less than its lower limit is defined as  
unity. Note that in this expression we have only one
complex integration for each complex conjugate eigenvalue pair. 
{The differentials of the complex eigenvalues are defined over the
  real and imaginary part, i.e. $d^{\,2}z_m=d\re z_m d\im z_m$.} 
In the following we treat the cases of
an even $(\chi=0)$ and odd $(\chi=1)$ total number of eigenvalues on
the same footing.
The jpdf for a fixed number $K$ of real eigenvalues and $M$
complex eigenvalue pairs is defined as  
\begin{eqnarray}
 \fl
P_{K,M}(x,z,z^*)&\equiv&
\prod\limits_{k=1}^{K}h(x_k)
\prod\limits_{m=1}^M\Big(g(z_m,z^*_m)\,2i\Theta(\im\,z_m)\Big)
\ \Delta_{K+2M}(\{x\},\{z,z^*\})\nonumber\\
\fl&&\times\prod\limits_{k=2}^{K}\Theta(x_k-x_{k-1})
\prod\limits_{m=2}^{M} 
\Theta(\re\,z_{m}-\re\,z_{m-1})
\label{jpdfPKM}
\end{eqnarray}
with the weight specified in eqs.\ (\ref{wch}) and (\ref{wgin}) below
\footnote{In contrast to \cite{APSoII} we distinguish the weights for real and
complex eigenvalues by different symbols ($h$ and $g$ respectively).}. 
Here $\Theta$ is the Heaviside distribution, and
the Vandermonde determinant is defined as
\be
\Delta_{N}(\{z\})=\prod_{k>l}^N(z_k-z_l)= \det_{1\leq a,b\leq
  N}\Big[z_a^{b-1}\Big]\ .
\label{Vandermonde}
\ee
In eq.\ (\ref{jpdfPKM}) we explicitly specify the number $K$ of real 
eigenvalues and $M$ complex eigenvalue pairs, with the set of
arguments labelled as $x_1,\ldots,x_K,z_1,z_1^*,\ldots,z_M,z_M^*$ in
$\Delta_{K+2M}(\{x\},\{z,z^*\})$.
The factors $2i\Theta(\im\,z_m)$ and the ordering of the real eigenvalues
$\Theta(x_k-x_{k-1})$ in eq.\ (\ref{jpdfPKM})
allow us to omit the modulus sign
around the Vandermonde determinant. 
The ordering of the
real parts $\Theta(\re\,z_{m}-\re\,z_{m-1})$ is needed to make the
transformation to upper triangular {$2\times2$ block} 
form of the matrices $A$ and $B^T$ (or $J$) unique when
computing the Jacobian \cite{APSoII} (see also \cite{Sommers2008}). 
The latter and part of the former can be dropped later due to the
symmetrising integration as will be shown in the next section.

We also mention that the partition function can be written as a single
Pfaffian \cite{Sinclair,Sommers2007}, eq.\ (\ref{ZPf}) below, 
and we come back to the consequences for factorisation in the next section.

We now give the weight functions for our two models. Looking at eq.\
(\ref{evrel}) 
for the chiral model 
we are only interested in the eigenvalues of the matrix
$C=AB^T$. Because the $N_f$ extra mass terms compared with \cite{APSoII}
depend only on $C$ their addition to the jpdf in \cite{APSoII} is
trivial and so we only give the result. 
The corresponding
weight functions in eq.\ (\ref{jpdfPKM}) 
for the real eigenvalues $x$ and complex eigenvalues
$z=x+iy$ read
\bea
\label{wch}
\fl h_{ch}(x) &\equiv&
2|x|^{\nu/2}K_{{\nu}/{2}}(\eta_+|x|)\exp[{\eta_-x}]
\prod_{f=1}^{N_f}(x+m_f^2)
\ ,\\
\fl g_{ch}(z_1,z_2)&\equiv&
2|z_1z_2|^{\nu/2}\exp[{\eta_-(z_1+z_2)}]
\prod_{f=1}^{N_f}(z_1+m_f^2)(z_2+m_f^2)
\nn\\
\fl&\times&
\int_0^\infty \frac{dt}{t}\exp\Big[-\eta_+^2 t(z_1^2+z_2^2)-\frac{1}{4t}\Big]
K_{{\nu}/{2}}\left(2\eta_+^2t z_1 z_2\right)
\erfc\left(\eta_+\sqrt{t}|z_2-z_1|\right).
\nn
\eea
In addition we have a trivial overall factor arising from the generic
zero eigenvalues 
in the mass terms:
\be
P_{K,M}^{ch}(x,z,z^*)\ =\prod_{f=1}^{N_f}m_f^\nu\ P_{K,M}(x,z,z^*)
\label{jpdfch}
\ee

For the second model eq.\ (\ref{Zgin}) we have instead
\bea
\label{wgin}
h_{Gin}(x) &\equiv&
\exp[{-x^2}]\prod_{f=1}^{N_f}(x^2+m_f^2)
\ ,\\
g_{Gin}(z_1,z_2)&\equiv&
\exp[{-z_1^2-z_2^{2}}]\ \erfc\left(\frac{|z_1-z_2|}{\sqrt{1-\tau}}\right)
\prod_{f=1}^{N_f}(z_1^2+m_f^2)(z_2^2+m_f^2)\ .
\nn
\eea
For the two Gaussian examples above the following relation is satisfied
\be
\lim_{\Im m \, z\to0}g(z,z^*)=h(\re \, z)^2\ ,
\label{hg}
\ee
relating the two weights.

As a general remark here and in the following we can allow for
more general weight functions $h(x)$ and $g(z_1,z_2)$
in eq.\ (\ref{jpdfPKM}) 
that do not necessarily
follow from a matrix representation. 
For example,
we could generalise the weights in eqs.\ (\ref{wch}) and (\ref{wgin})
by multiplying by a factor $\exp[-V(z_1,z_2)]$
where $V$ is a polynomial in $z_1$ and $z_2$ \footnote{A so-called
  harmonic potential could be realised as a matrix model, by
  multiplying eqs.\ (\ref{Zch}) and (\ref{Zgin}) with $\exp[-\Tr
    V(AB^T)]$ or $\exp[-\Tr V(J^2)]$. 
Although at finite $N$ these are formal expressions due to lack of
convergence this can be dealt with in the large-$N$ limit.
}. Moreover one can independently choose $h$ and $g$ instead
fulfilling the relation \eref{hg}.

\sect{Factorisation of the joint probability distribution}\label{jpdf}

In this section we prove that 
the jpdf inside the partition function can be written in a factorised form. 
For this to be possible, it is essential that we integrate over
all the eigenvalues, leading to a symmetrisation.
Hence this applies equally to the expectation value of
any operator symmetric in all variables. Examples for this are the
computation of the gap probability or expectation values
leading to integral representations of the SOP in the next section. 
However, such a factorisation can also be found for 
the $k$-point density correlation functions when summing over all
possibilities of splitting $k$ into real and complex eigenvalue pairs.

Let us first state the result for the partition function eq.\ (\ref{Zsum})
in terms of a single product for the weights
\be
\fl {\cal Z}_{2N+\chi}\ =\ 
\int\limits_{\mathbb{R}}dy^\chi\ h^\chi(y)
\prod_{k=1}^{2N}  \int\limits_{\mathbb{C}} d^{\,2}z_k
\prod\limits_{j=1}^{N}F(z_{2j-1},z_{2j})
\ \Delta_{\chi+2N}(y,\{z\})\ ,
\label{Zgen}
\ee
where we define the anti-symmetric function
\begin{eqnarray}
\label{Fdef}
\fl F(z_{1},z_{2})
&\equiv&
{i}g(z_{1},z_{2})(\Theta(\im\,z_{1})-\Theta(\im\,z_{2}))\,
\delta^2(z_{2}-z_{1}^*)\\
\fl &&+  \frac{1}{2}  h(z_{1})h(z_{2})\delta(\im\,z_{1})\delta(\im\,z_{2})
\sgn(\re\,z_{2}-\re\,z_{1})\, .\nn
\end{eqnarray}
Note that for an even number of variables
($\chi=0$ in eq.\ (\ref{Zgen})) the integration and weight
for the real variable $y$ have to be dropped, as well as the argument
$y$ inside the Vandermonde determinant.
In  eq.\ (\ref{Zgen}) we 
integrate over $2N$ independent complex
variables in contrast to the 
ordered integration in complex conjugated pairs.
The standard 
two-dimensional delta function in eq.\ (\ref{Fdef}) reads 
$\delta^2(z)=\delta(x)\delta(y)$ for $z=x+iy$.

We will prove this factorisation in two different ways. One is by explicitly
summing up all the terms in eq.\ (\ref{Zsum}) to make a single
product. {The results for expectation values of characteristic polynomials 
in \cite{KGII} are built up on this idea, although 
a proof of this was not given.} 
The
second way is by starting from a single Pfaffian representation of the
partition function derived in \cite{Sinclair,Sommers2007} and using
(in reverse) a proof of a slightly 
generalised version of the de Bruijn integral formula. Because both
derivations are short and illustrate different aspects we decided to
present both.

The first derivation of the factorisation of the jpdf
goes as follows. 
From eqs.\ (\ref{jpdfPKM}) and (\ref{Zsum}) we obtain
\begin{eqnarray}
 \fl 
{{\cal Z}_{2N+\chi}}&=&
N!\sum\limits_{n=0}^{N}\frac{1}{(N-n)!}\
\prod_{l=1}^{2n+\chi}\int\limits_{\mathbb{R}} dx_l\,h(x_l)
\prod\limits_{j=2}^{2n+\chi}\Theta(x_j-x_{j-1})\nonumber\\
\fl&&\times
\prod_{m=1}^{N-n}\int\limits_{\mathbb{C}}(d^{\,2}z_m \,
g(z_m,z^*_m)\,2i\Theta(\im\,z_m))
\ \Delta_{(2n+\chi)+2(N-n)}(\{x\},\{z,z^*\})\ ,
\label{symm1}
\end{eqnarray}
where we use that 
the integrand 
is totally symmetric under a permutation of two pairs of complex
conjugated eigenvalues.
Dropping
the ordering of the real parts 
leads to a factor $1/{(N-n)!}$.

The ordering of the real variables can be simplified by applying the method of
integration over alternating variables \cite{Mehta} twice. 
Pulling the integrations over odd variables in, dropping the
symmetrisation giving a factor $1/n!$ and then pulling them
back out leads to the following result:
%
\bea
\fl
{{\cal Z}_{2N+\chi}}&=&
\sum\limits_{n=0}^{N}\frac{N!}{n!(N-n)!}
\prod_{l=1}^{2n+\chi}\int\limits_{\mathbb{R}}dx_l\,h(x_l)
\prod\limits_{j=1}^{n}\Theta(x_{2j}-x_{2j-1})\nn\\
\fl&&\times
\prod_{m=1}^{N-n}\int\limits_{\mathbb{C}}(d^{\,2}z_m \,
g(z_m,z^*_m)\,2i\Theta(\im\,z_m))
\ \Delta_{(2n+\chi)+2(N-n)}(\{x\},\{z,z^*\})\ .
\eea
Isolating the integration $dx_\chi$ we can do the sum over multiple
integrations, using the
binomial formula
and the permutation invariance of the
integrand  under exchanging pairs of complex numbers:
\bea
\fl a&\equiv& \int\limits_{\mathbb{R}}dx_1\,h(x_1) 
\int\limits_{\mathbb{R}}dx_2\,h(x_2)\,\Theta(x_{2}-x_{1})\nn\\
\fl &=&
\int\limits_{\mathbb{C}}d^{\,2}z_1\, h(z_1)
\int\limits_{\mathbb{C}}d^{\,2}z_2 \,h(z_2)\,\Theta(\re\,z_{2}-\re\,z_{1})
\,\delta(\im\,z_{1})\,\delta(\im\,z_{2})\\
\fl b&\equiv&\int\limits_{\mathbb{C}}d^{\,2}z \,
g(z,z^*)\,2i\Theta(\im\,z)
=\int\limits_{\mathbb{C}}d^{\,2}z_1 
\int\limits_{\mathbb{C}}d^{\,2}z_2 \,
g(z_1,z_2)\,2i\Theta(\im\,z_1)\,\delta^2(z_{2}-z_{1}^*)\ ,
\eea
with
\be
\sum\limits_{n=0}^{N}\frac{N!}{n!(N-n)!} a^n b^{N-n}\ =\ (a+b)^N\ .
\ee
Pulling out all $2N$ independent complex integrations of the $N$-fold
product this leads to eq.\ (\ref{Zgen}) as claimed, 
after making the function $F(z_1,z_2)$ manifestly anti-symmetric.


We now come to our second argument, starting from the result 
derived in \cite{Sinclair,Sommers2007}. This states that, 
including normalisation,
\be
{\cal Z}_{2N}\ =\ N!\,
\underset{1\leq a,b\leq 2N}{\Pf}\left[ \,\, 
\int\limits_{\mathbb{C}^2}d^{\,2}z_1d^{\,2}z_2 
F(z_1,z_2)[z_1^{a-1}z_2^{b-1}-z_2^{a-1}z_1^{b-1}]
\right],
\label{ZPf}
\ee
where for simplicity we only state the even case, i.e.\ with $2N$
eigenvalues. {The sign of the Pfaffian is defined as in
  \cite{Mehta} such that the Pfaffian of the} {matrix
  $\left[\begin{array}{cc} 0 & 1 \\ -1 & 0 \end{array}\right]\otimes
  \eins_N$ is unity.} 
Here $F(z_1,z_2)$ is the function from eq.\ (\ref{Fdef}).
By using a slight generalisation of the proof of the de Bruijn
  integral formula, in reverse,
\bea
\fl&&\prod_{k=1}^{2N}\int\limits_{\mathbb{C}} d^{\,2}z_k
\prod\limits_{j=1}^{N}F(z_{2j-1},z_{2j})
\det_{1\leq a,b\leq N}
[\{f_a(z_{2b-1}),g_a(z_{2b})\}]\nn\\
\fl&=&N!\ \underset{1\leq a,b\leq 2N}{\Pf}\left[
\int\limits_{\mathbb{C}^2}
d^{\,2}u \, d^{\,2}vF(u,v)[f_a(u)g_b(v)-f_b(u)g_a(v)]
\right]
\label{deB}
\eea
where we refer to \ref{GendB} for a derivation (cf. 
Appendix C.2 in \cite{KGI}), we obtain
\be
{\cal Z}_{2N}\ =
\prod_{k=1}^{2N}\int\limits_{\mathbb{C}} d^{\,2}z_k
\prod\limits_{j=1}^{N}F(z_{2j-1},z_{2j})
\det_{1\leq a\leq 2N;\,1\leq b\leq N}
\left[\{z_{2b-1}^{a-1},z_{2b}^{a-1}\}\right].
\ee
Here the last determinant is simply the Vandermonde determinant, and
thus we have arrived again at eq.\ (\ref{Zgen}). 

All of the above arguments can be generalised, 
by including an arbitrary observable that is symmetric under the
exchange of all the eigenvalues. Perhaps the simplest example, which we will
also encounter in the next section, is a
factorising operator,
\be
f(x,z,z^*)=f^{\chi}(x) \prod\limits_{j=1}^{2N}f(z_j)\,.
\ee
The individual factors do not affect the symmetry arguments above and
we obtain
\bea
\fl&&\langle f(x,z,z^*) \rangle_{2N+\chi} = \frac{1}{{\cal Z}_{2N+\chi}N!}
\int\limits_{\mathbb{R}}\!dy^\chi(h(y)f(y))^\chi
\prod_{k=1}^{2N} \int\limits_{\mathbb{C}} d^{\,2}z_kf(z_k)
\prod\limits_{j=1}^{N}F(z_{2j-1},z_{2j})
\nn\\
\fl&&\qquad\qquad\qquad\qquad\qquad \qquad\qquad
\times \Delta_{\chi+2N}(y,\{z\}) 
\label{vevfact}
\eea
for general expectation values.
An explicit example for such an operator is the characteristic
polynomial. 
{This result can be generalised to
  non-factorising observables symmetric in the eigenvalues. Since the
  monomials in the traces of a matrix can be traced back to products
  over characteristic polynomials eq.~\eref{vevfact} is true for all
  symmetric polynomials in the eigenvalues. With Weierstra\ss'
  approximation theorem all symmetric functions are built of
  polynomials in the traces, and as a limiting case
one can also construct
  distributions like the Dirac distribution. This means that all
  observables symmetric in the eigenvalues fulfil a similar equation
  as \eref{vevfact} in a weak sense.}

As a further remark the symplectic non-Hermitian ensembles with $\beta=4$
\cite{A05,Kan02} are already of the factorised form eq.\ (\ref{Zgen})
from the onset, with 
\be
F^{(\beta=4)}(z_1,z_2)=i(z_2-z_1)w(z_1,z_2)\delta^2(z_{2}-z_{1}^*)
\label{F4map}
\ee
and we always have $\chi=0$ due to symmetry. The symmetric real weight 
$w(z_1,z_2)$ can be found in \cite{Kan02} and \cite{A05} for the 
Gaussian Ginibre and chiral classes respectively. 
Therefore all statements we derive from the form of 
eq.\ (\ref{Zgen}) automatically hold true for these symmetry classes
as well.  The factorisation thus unifies the non-Hermitian ensembles for
$\beta=1$ and $4$; in fact, this  was already pointed out in 
{\cite{BHJ,KGII}. In \cite{BHJ}, this was found} in a different way
without using factorisation. 

\sect{Integral representation 
of skew-orthogonal polynomials and their Cauchy transforms
}\label{intrep}

In this section we will derive integral representations for the SOP
for general weight functions, using the results from the previous
section. For this we will only need the result for an even total number of
eigenvalues (i.e.\ $\chi=0$).

All the matrix or complex eigenvalue models introduced previously 
can be solved for all eigenvalue density correlation functions in
terms of the following skew-symmetric kernel
\begin{eqnarray}
\mcK_{2N}(z_1,z_2)&=& \sum_{k,l=0}^{2N-1}
{\mcA}_{kl}^{-1} \, p_k(z_1) \, p_l(z_2)
\label{Kdef}
\end{eqnarray}
where
\bea
{\mcA}_{kl} & \equiv & 
{2}\int\limits_{\mathbb{C}^2} d^{\,2}z_1 \, d^{\,2}z_2 \,
F(z_{1},z_{2})\,
p_k(z_1) \, p_l(z_2)\ .
\eea
In fact, the kernel is
only a property of the measure $F(z_{2j-1},z_{2j})$ and not of 
the particular choice of the polynomials $\{p_k(z)\}$; in \cite{Sommers2007}
these were chosen to be monic.
For an odd total number of eigenvalues a similar
representation to eq.\ (\ref{Kdef}) holds, but with a modification to
the last row and 
column of the matrix ${\mcA}$; see \cite{Sommers2008} and
\cite{ForresterMays}. 
Here we will choose the polynomials $p_k(z)$ to be skew-orthogonal with respect
to the following anti-symmetric scalar product 
\begin{equation}\label{skewdef}
 \langle f|g\rangle=-
\langle g|f\rangle\equiv
\int\limits_{\mathbb{C}^2}d^{\,2}z_1\,d^{\,2}z_2\, 
F(z_1,z_2)\det\left[\begin{array}{cc}
    f(z_1) & g(z_1) \\ f(z_2) & g(z_2) \end{array}\right],
\end{equation}
defined for two functions $f(z)$ and $g(z)$ that are integrable with
respect to the weight functions contained in $F(z_1,z_2)$.
This includes the particular function $g(z_1,z_2)$ from eq.\ (\ref{jpdfPKM}).

Our skew-orthogonal polynomials $q_k(z)$ are defined to satisfy
\bea
\langle q_{2k}|q_{2l+1}\rangle&=&h_k\delta_{kl}\ ,\nn\\
\langle q_{2k}|q_{2l}\rangle&=&0\ =\ \langle q_{2k+1}|q_{2l+1}\rangle
\ \forall k,l\geq0\ ,
\label{qdef}
\eea
where the $h_k>0$ are their positive (squared skew) norms, see
eq. (\ref{Zprod}).   
This leads to a block diagonal matrix
$\mcA=\diag(h_0\epsilon,\ldots,h_{N-1}\epsilon)$ that can be easily inverted,
where $\epsilon$ is the anti-symmetric $2\times 2$ matrix with elements 
$\epsilon_{12}=1=-\epsilon_{21}$. 
The kernel can be written as a single sum in terms of the
SOP:
\be
\label{kernel_sum_qq}
\mcK_{2N}(z_1,z_2) = \sum_{k=0}^{N-1} \, \frac{1}{h_k} \, \big(
q_{2k+1}(z_1)q_{2k}(z_2) - q_{2k+1}(z_2)q_{2k}(z_1) \big).
\ee
{The
kernel for $2N+1$ contains the same SOP plus a correction term, see
also \cite{ForresterMays} 
for the Ginibre ensemble and \cite{Mehta} for the GOE.} 

\subsection{Skew-orthogonal polynomials}\label{sops}

After all this preparation we come to our second result, an explicit
representation of the SOP. They are given in terms of the following
expectation values
\begin{equation}\label{idq1}
\fl q_{2n}({z})=\Big\langle
 \det({z}-J)\Big\rangle_{2n}
=\left\langle
 \prod\limits_{j=1}^{2n}(z-z_j)\right\rangle_{2n}
\end{equation}
for the even polynomials, and
\begin{eqnarray}
\fl q_{2n+1}({z})&=&\Big\langle \det({z}-J)[\Tr J+z+c]\Big\rangle_{2n}
\ =\ \Big\langle \det({z}-J)\Tr J\Big\rangle_{2n}\ +\ (z+c)\, q_{2n}({z})
\nn\\
\fl&=&\left\langle\prod\limits_{j=1}^{2n}(z-z_j)
\Big[\sum\limits_{i=1}^{2n}z_i\ +\ z\ +\ c\Big]\right\rangle_{2n}
\,,\label{idq2}
\end{eqnarray}
for the odd polynomials, which are both expectation values over an
even number of eigenvalues $2n$, $n\geq1$. 
For $n=0$ we simply have $q_0(z)=1$ and $q_1(z)=z+c$, by definition.
It is easy to see by taking large arguments that these representations
are in monic normalisation, {\it viz} $q_n(z)=z^n+{\cal O}(z^{n-1})$.
Similar expressions hold in terms of the matrix $\mathcal{D}$ from
eq.\ (\ref{Mdef}) for the chiral model 
(see subsection \ref{sLaguerre} for more details), whilst the
representations given in terms of squared eigenvalues $\La_j^2=z_j$ are
identical. Eqs.\ (\ref{idq1}) and (\ref{idq2}) were also shown for
particular non-Hermitian ensembles in refs. \cite{Eynard,GP}.

The set of odd polynomials is not unique because of the anti-symmetry of
the skew product (\ref{skewdef}), as the arbitrary constant $c$ times the
even polynomial drops out. In most of the following we will keep the
constant $c\neq0$ though.

The proof of the first integral representation eq.\ (\ref{idq1})
goes as follows. We can write the
product times the Vandermonde determinant of dimension $2n$ as a
Vandermonde determinant of 
dimension $2n+1$, and so
\be
\fl q_{2n}({z})=\frac{1}{{\cal Z}_{2n}}
\prod_{k=1}^{2n} \int\limits_{\mathbb{C}} d^{\,2}z_k
\prod\limits_{j=1}^{n}F(z_{2j-1},z_{2j})
\det_{1\leq a\leq 2n+1;1\leq b\leq 2n}
\left[\{z_b^{a-1}\}|z^{a-1}\right].
\label{preq2n}
\ee

We can now apply a slight modification of the generalisation of the 
de Bruijn integral formula
proved in {Appendix C.2} of \cite{KGI},
\bea
\fl&&\prod_{k=1}^{2n} \int\limits_{\mathbb{C}} d^{\,2}z_k
\prod\limits_{j=1}^{n}F(z_{2j-1},z_{2j})
\det_{1\leq a\leq 2n+m;\,1\leq j\leq n;\,1\leq i\leq m}
\Big[\{f_a(z_{2j-1}),g_a(z_{2j})\}|\ \al_{ai}\ \Big]\nn\\
\fl=&&
(-)^{m(m-1)/2}n!\underset{1\leq a,b\leq
  2n+m;\,1\leq i\leq m}{\Pf}\left[\!\!
\begin{array}{ll}
\left\{\int\limits_{\mathbb{C}^2}d^{\,2}u \, d^{\,2}v \,
F(u,v)[f_a(u)g_b(v)-f_b(u)g_a(v)] 
\right\}\!&\!\al_{ai}\\
-\al^T_{ib}&\!0\\
\end{array}\!
\right]
\nn\\
\fl&&
\label{gendeB}
\eea
The overall sign can be seen by choosing
$[\al_{ai}]=\left[\ \mathbf{0}_{m,\,2n}\ \eins_m\ \right]^T$.
Here $\al$ is a constant matrix, which in our case in eq.\ (\ref{preq2n})
is a simple vector with $m=1$. In contrast to the usual de Bruijn
formula 
we integrate over $2n$ variables here instead of $n$, as is shown to
hold in \ref{GendB} (see also Appendix C.2 in \cite{KGI}).

Denoting the basis functions of monic powers by $e_a(z)\equiv z^a$ and
using the fact 
that we have equal functions $f_a(z)=g_a(z)=e_{a-1}(z)$ above we arrive at
\be
q_{2n}({z})=\frac{n!}{{\cal Z}_{2n}}
\underset{1\leq a,b\leq 2n+1}{\Pf}\left[
\begin{array}{ll}
\{\langle e_{a-1}|e_{b-1}\rangle\} &e_{a-1}(z)\\
-e_{b-1}(z)&0\\
\end{array}
\right].
\label{q2n}
\ee
It is easy to see that using the definition of the skew product in
eq.\ (\ref{skewdef}) 
and performing one more integration we have 
\be
\langle q_{2n}|e_{c}\rangle\ =\ 0\ \ \forall c=0,\ldots 2n\ ,
\label{qeON}
\ee
because the corresponding Pfaffian vanishes.
Using the linearity of the skew product 
we can deduce that the even polynomials $q_{2n}(z)$ in
eq. (\ref{idq1}) are skew-orthogonal to all
polynomials of lower and equal degree.

To prove the second integral representation eq.\ (\ref{idq2})
we need a further identity for manipulating Vandermonde determinants,
\be
\sum_{a=1}^{N}z_a\ \Delta_{N}(\{z\})\ =\ \det\left[
\begin{array}{lll}
1  &\ldots&1\\
z_1&\ldots&z_{N}\\
\vdots&&\vdots\\
z_1^{N-2}&\ldots&z_{N}^{N-2}\\
z_1^{N}&\ldots&z_{N}^{N}\\
\end{array}
\right]
\equiv \widetilde{\Delta}_{N}(\{z\})
\ ,
\label{Vid}
\ee
which is proved in \ref{VIdI}. We can
now proceed as in eq.\ (\ref{preq2n}), by first incorporating the product in
eq.\ (\ref{idq2}) into a larger Vandermonde determinant, and then applying
the identity (\ref{Vid}) for $2n+1$: 
\be
\fl q_{2n+1}({z})=\frac{1}{{\cal Z}_{2n}}
\prod_{k=1}^{2n} \int\limits_{\mathbb{C}} d^{\,2}z_k
\prod\limits_{j=1}^{n}F(z_{2j-1},z_{2j})
\det_{1\leq a,b\leq 2n}
\left[
\begin{array}{cc}
\{z_b^{a-1}\}&z^{a-1}\\
z_b^{2n+1}&z^{2n+1}\\
\end{array}
\right].
\label{preq2n+1}
\ee
For simplicity we have set $c=0$ here as it does not affect the proof.
Again applying the integral formula eq.\ (\ref{gendeB}), with a slightly
modified range of indices compared with the even polynomial case, we obtain
\be
\fl q_{2n+1}({z})=\frac{n!}{{\cal Z}_{2n}}
\underset{1\leq a,b\leq 2n+1}{\Pf}\left[
\begin{array}{ccc}
\{\langle e_{a-1}|e_{b-1}\rangle\}
&\langle e_{a-1}|e_{2n+1}\rangle &e_{a-1}(z)\\
\langle e_{2n+1}|e_{b-1}\rangle &0 &e_{2n+1}(z)\\
-e_{b-1}(z)&-e_{2n+1}(z)&0\\
\end{array}
\right].
\label{q2n+1}
\ee
On taking the skew product (\ref{skewdef}) of this result it obviously
follows that 
\be
\langle q_{2n+1}|e_{c}\rangle\ =\ 0\ \ \forall c=0,\ldots 2n-1\ ,
\label{qoON}
\ee
this being the skew-orthogonality of the odd polynomials $q_{2n+1}(z)$
given by eq. (\ref{idq2}) 
to all polynomials of degree less than or equal to $2n-1$.

As a last step we will verify the coefficient of the only non-vanishing
skew product which due to linearity and eq.\ (\ref{qeON}) equals
$\langle q_{2n}|e_{2n+1}\rangle=\langle
q_{2n}|q_{2n+1}\rangle$. 
To do so we will first determine the partition function in terms of
the norms. 
It follows along the lines of
eq.\ (\ref{preq2n}). Inside the Vandermonde determinant there 
we could choose any
set of polynomials in monic normalisation, after applying the
invariance properties of the determinant. We thus have 
\bea
\fl{{\cal Z}_{2n}}&=&
\prod_{k=1}^{2n} \int\limits_{\mathbb{C}} d^{\,2}z_k
\prod\limits_{j=1}^{n}F(z_{2j-1},z_{2j})
\det_{1\leq a,b\leq 2n}
\left[q_{a-1}(z_b)\right]
\ =\ n!\ \underset{1\leq a,b\leq 2n}{\Pf}[\langle q_{a-1}|q_{b-1}\rangle]\nn\\
\fl&=& n!\prod_{a=0}^{n-1}h_a\ .
\label{Zprod}
\eea
In the second step we applied once more the integral identity
(\ref{gendeB}), with $m=0$ and the matrix $\al$ absent. Due to the
skew-orthogonality, the matrix inside the Pfaffian becomes block
diagonal, with the norms $h_k$ times $\epsilon$ down the diagonal, which
finally leads to the product of the norms.

Using eq.\ (\ref{q2n}) after replacing the monic powers with
polynomials $q_k$ we have 
\bea
\fl\langle q_{2n}|q_{2n+1}\rangle &=&\frac{n!}{{\cal Z}_{2n}}
\underset{1\leq a,b\leq 2n+1}{\Pf}\left[\begin{array}{cc}
\{\langle q_{a-1}|q_{b-1}\rangle\} &\langle q_{a-1}|q_{2n+1}\rangle \\
-\langle q_{2n+1}|q_{b-1}\rangle&0\\
\end{array}
\right]\nn\\
\fl &=&\frac{n!}{{\cal Z}_{2n}}\prod_{a=0}^{n}h_a\ =\ h_n\ ,
\eea
and thus the consistency of the normalisation of our
integral representations (\ref{idq1}) and (\ref{idq2}) with respect to 
eq.\ (\ref{qdef}).
This concludes our proof of the integral representations of the SOP
satisfying eq.\ (\ref{qdef}).
An entirely different derivation of the same results can be made by
a mapping to the $\beta=4$ symplectic case. When mapping our
$F(z_1,z_2)$ as in eq.\ (\ref{F4map}) we could in principle copy the
orthogonality proof from \cite{Kan02} where the representations eq.\
(\ref{idq1}) 
and (\ref{idq2}) were derived for $\beta=4$ .

It is worth mentioning that the same representation for SOP 
holds for Hermitian RMT at $\beta=1$ and $4$
with real eigenvalues as was shown earlier in \cite{Eynard,GP}. 
However, for all four cases -- two Hermitian and two non-Hermitian --
the jpdf and corresponding skew products are different. 
In contrast for $\beta=2$ all OP are obtained from a single relation as in
eq.\ (\ref{idq1}), in both Hermitian and non-Hermitian RMT
\cite{AV03}.
The fact that the same integral representation for SOP holds both in non-chiral
\cite{Eynard,GP} and chiral ensembles is straightforward in the
Hermitian case. However, for non-Hermitian ensembles this becomes 
nontrivial comparing \cite{Kan02} vs. \cite{A05} for $\beta=4$, and \cite{APSo}
for $\beta=1$. This is due to the two-matrix model structure of the chiral
ensembles, where the change to an eigenvalue basis requires detailed 
calculations. 

Let us finish this subsection with some remarks.
For an even number of eigenvalues $\chi=0$
the anti-symmetric kernel eq.\ (\ref{Kdef}) (or (\ref{kernel_sum_qq})) 
can itself be expressed as
the expectation value of two characteristic polynomials for $\beta=1$
\cite{APSo}\footnote{Note that the overall constant in front of the
  kernel has been chosen here to be consistent with the standard
  choice in eq.\ (\ref{kernel_sum_qq}).} (and $\beta=4$ \cite{AB})
\be
\Big\langle \det (\lambda-J) \det(\gamma-J) \Big\rangle _{2N}
\ =\ h_N\frac{{\cal K }_{2N+2} (\lambda ,\gamma)}{\lambda-\gamma}\ \
\mbox{with}\ \ \lambda\neq\gamma\ ,
\label{idkernel}
\ee
and similarly for the chiral ensemble.
This equation is valid for arbitrary weight functions.
In fact we will partly use this relation to determine the set
of odd polynomials eq.\ (\ref{idq2}) in section \ref{examples} below.
So why are eqs.\ (\ref{idq1}) and (\ref{idq2}) interesting if the
kernel itself can be independently determined as a building block?
It is because the integral representations we just derived, 
and the explicit determination of the SOP in some examples in the
next section,
complete the list of classical polynomials in the complex plane for
the three elliptic Ginibre ensembles and their chiral extensions.

The determination of the SOP through an ansatz, and subsequently the
direct verification of the relations (\ref{qdef}) for skew-orthogonal
Hermite polynomials, was already a
formidable task for the elliptic real Ginibre ensemble as can be seen from
\cite{Forrester07}. Because of the non-Gaussian form of the chiral
weight eq.\ (\ref{wch}) this is even more so true for skew-orthogonal
Laguerre polynomials. The integral representations derived here 
provide an alternative, constructive approach, leading to a new result
for skew-orthogonal Laguerre polynomials.

\subsection{Cauchy transforms}\label{cauchy}

We now come to the definition and integral representation of the
Cauchy transforms $t_k(z)$. It is very natural to define them with respect to
the scalar product eq.\ (\ref{skewdef}) as follows:
\be
\fl t_n(\kappa)\equiv 
\int\limits_{\mathbb{C}^2}d^{\,2}z_1\,d^{\,2}z_2\, 
F(z_1,z_2)\det\left[\begin{array}{cc}
    q_n(z_1) & \displaystyle\frac{1}{\ka-z_1} \\ q_n(z_2) &
    \displaystyle\frac{1}{\ka-z_2}  
\end{array}\right]
=\left\langle q_n \Big| \frac{1}{\ka-z} \right\rangle\ .
\label{tndef}
\ee
We will now show that the following integral representations hold:
\be
\label{idt1}
\fl t_{2n}({\ka})=h_n\left\langle
\frac{1}{\det({\ka}-J)}\right\rangle_{2n+2}
=h_n\left\langle
 \prod\limits_{j=1}^{2n+2}\frac{1}{(\ka-z_j)}\right\rangle_{2n+2}
\ee
for the Cauchy transforms of the even polynomials, and
\bea
\fl t_{2n+1}({\ka})&=&h_n\left\langle \frac{\Tr J-(\ka+c)}{\det({\ka}-J)}
\right\rangle_{2n+2}
\ =\ h_n\left\langle \frac{\Tr J}{\det({\ka}-J)}\right\rangle_{2n+2}\ -\
(\ka+c)\, t_{2n}({\ka})
\nn\\
\fl&=&h_n\left\langle
\frac{\sum\limits_{a=1}^{2n+2}z_a\ -(\ka+c)}{
\prod\limits_{j=1}^{2n+2}(\ka-z_j)}\right\rangle_{2n+2}
\, \label{idt2}
\eea
for the odd polynomials.
Note that the averages for $t_{2n}({\ka})$ and $t_{2n+1}({\ka})$
run over $2n+2$ variables, instead of $2n$ as for the polynomials
$q_{2n}(z)$ and $q_{2n+1}(z)$. This implies in particular that
$t_{0}(\ka)\neq$ constant, see also eq.\ (\ref{texpand}) below.

The correct overall prefactors can also easily
be seen.
From expanding the geometric series in the definition
eq.\ (\ref{tndef}) for large arguments, and using eqs.\ (\ref{qeON})
and (\ref{qoON}) as well as the anti-symmetry of the first
non-vanishing skew product, it follows that the Cauchy 
transforms are indeed 
Laurent series with the following coefficients 
\bea
t_{2n}(\ka)&=& +\ \frac{h_n}{\ka^{2n+2}}\ +\ 
{\cal O}\left( \frac{1}{\ka^{2n+3}}\right), \nn\\
t_{2n+1}(\ka)&=& -\ \frac{h_n}{\ka^{2n+1}}\ +\ 
{\cal O}\left( \frac{1}{\ka^{2n+2}}\right). 
\label{texpand}
\eea

The form given in eqs.\ (\ref{idt1}) and (\ref{idt2})
is completely analogous to eqs.\ (\ref{idq1}) and (\ref{idq2}), as
well as to the corresponding result for $\beta=2$. Let us also remark
that such a representation was not known before in the non-Hermitian $\beta=4$
symmetry class, and that both translate into new representations for
$\beta=1,4$ in the Hermitian limit.

We begin by proving the representation for the Cauchy transforms of the even
polynomials. Inserting the result eq.\ (\ref{idq1})
into the definition {(\ref{tndef})} we have 
\bea
\fl t_{2n}(\kappa)&=&
\int\limits_{\mathbb{C}^2}d^{\,2}z_{1}\,d^{\,2}z_{2}\, 
F(z_{1},z_{2})\left[
\frac{\langle\det(z_1-J) \rangle_{2n}}{\ka-z_2}
-\frac{\langle\det(z_2-J) \rangle_{2n}}{\ka-z_1}  \right] 
\nn\\
\fl &=&\frac{1}{{\cal Z}_{2n}}
\int\limits_{\mathbb{C}^2}
d^{\,2}z_{2n+1}\,d^{\,2}z_{2n+2}\, 
F(z_{2n+1},z_{2n+2})
\prod_{k=1}^{2n} \int\limits_{\mathbb{C}} d^{\,2}z_k
\prod\limits_{j=1}^{n}F(z_{2j-1},z_{2j}) \ \Delta_{2n}(\{z\})\nn\\
\fl &&\times\left[
\frac{\prod_{j=1}^{2n}(z_{2n+1}-z_j) }{\ka-z_{2n+2}}
-\frac{\prod_{j=1}^{2n}(z_{2n+2}-z_j) }{\ka-z_{2n+1}}
\right] 
\nn\\
\fl &=& \frac{1}{(n+1){\cal Z}_{2n}}
\prod_{k=1}^{2n+2} \int\limits_{\mathbb{C}} d^{\,2}z_k
\prod\limits_{j=1}^{n+1}F(z_{2j-1},z_{2j})
\nn\\
\fl &&\times
\sum_{k=1}^{n+1}\left[ \frac{\Delta_{2n+1}(\{z\}_{l\neq 2k})}{\ka-z_{2k}}
-\frac{\Delta_{2n+1}(\{z\}_{l\neq 2k-1})}{\ka-z_{2k-1}}
\right]
\nn\\
\fl &=& \frac{1}{(n+1){\cal Z}_{2n}}
\prod_{k=1}^{2n+2} \int\limits_{\mathbb{C}} d^{\,2}z_k
\prod\limits_{j=1}^{n+1}F(z_{2j-1},z_{2j})
\det_{1\leq a\leq 2n+2;\,1\leq b\leq 2n+1}\left[
\{z_a^{b-1}\}\Big|\frac{1}{\ka-z_a}\right]
\nn\\
\fl &=& \frac{1}{(n+1){\cal Z}_{2n}}
\prod_{k=1}^{2n+2} \int\limits_{\mathbb{C}} d^{\,2}z_k
\prod\limits_{j=1}^{n+1}F(z_{2j-1},z_{2j})\ 
\frac{\prod_{i>j}^{2n+2}(z_i-z_j)}{\prod_{l=1}^{2n+2}(\ka-z_l)}
\nn\\
\fl&=&h_{n}\left\langle
\frac{1}{\det(\ka-J)}\right\rangle_{2n+2}\ .
\label{idt1proof}
\eea
In the first step we have simply written out the expectation value and
renamed the additional two integration variables. The products in the
numerator can be incorporated into a larger Vandermonde determinant.
Next we can symmetrise
the integrand with respect to an exchange of any pair of variables
$z_{2j},z_{2j+1}$ leading to a prefactor $1/(n+1)$. The resulting
expression can be seen to be the expansion of a 
Vandermonde determinant plus an extra
column. In the last step we use an identity that was proved in
\cite{KGI}, see eqs.\ (3.3) vs (3.7) there\footnote{Note that we order
products here so that there is no sign in eq.\ (\ref{Vandermonde}) for
the Vandermonde determinant.},
deriving different representations for Berezinians
\be
\det_{1\leq a\leq 2n;\,1\leq b\leq 2n-1}\left[
\{z_a^{b-1}\}\Big|\frac{1}{\ka-z_a}
\right]\ =\
\frac{\prod_{a>b}^{2n}(z_a-z_b)}{\prod_{l=1}^{2n}(\ka-z_l)}\ .
\label{BidI}
\ee
This can be used to express the Cauchy transform as an expectation value,
after providing the correct normalisation factor from eq.\
(\ref{Zprod}) in the last step.

The proof for the odd Cauchy transforms follows along the same lines. For
simplicity we set $c=0$ here, which can easily be reinstated at the end:
\bea
\fl t_{2n+1}(\kappa)&=&\frac{1}{{\cal Z}_{2n}}
\prod_{k=1}^{2n+2} \int\limits_{\mathbb{C}} d^{\,2}z_k
\prod\limits_{j=1}^{n+1}F(z_{2j-1},z_{2j}) \left[
\frac{\Big(\sum_{l=1}^{2n}z_l\ +z_{2n+1}\Big)
\prod_{j=1}^{2n}(z_{2n+1}-z_j) }{\ka-z_{2n+2}}\right.
\nn\\
\fl &&
\left.
-\ \frac{\Big(\sum_{l=1}^{2n}z_l\ +z_{2n+2}\Big)
\prod_{j=1}^{2n}(z_{2n+2}-z_j) }{z_{2n+1}-\ka}
\right]\Delta_{2n}(\{z\}) \nn\\
\fl &=& \frac{1}{(n+1){\cal Z}_{2n}}
\prod_{k=1}^{2n+2} \int\limits_{\mathbb{C}} d^{\,2}z_k
\prod\limits_{j=1}^{n+1}F(z_{2j-1},z_{2j})
\nn\\
\fl &&\times
\sum_{k=1}^{n+1}\left[ \frac{\widetilde{\Delta}_{2n+1}(\{z\}_{l\neq 2k})
}{\ka-z_{2k}}
-\frac{\widetilde{\Delta}_{2n+1}(\{z\}_{l\neq 2k-1})}{\ka - z_{2k-1}}
\right]
\nn\\
\fl &=& \frac{1}{(n+1){\cal Z}_{2n}}\!
\prod_{k=1}^{2n+2} \int\limits_{\mathbb{C}}\!\! d^{\,2}z_k
\prod\limits_{j=1}^{n+1}F(z_{2j-1},z_{2j})\!
\det_{1\leq a\leq 2n+2;\,1\leq b\leq 2n}\!\left[
\{z_a^{b-1}\}\Big|z_a^{2n+1}\Big|\frac{1}{\ka-z_a}\right]
\nn\\
\fl &=& \frac{1}{(n+1){\cal Z}_{2n}}
\prod_{k=1}^{2n+2} \int\limits_{\mathbb{C}}\! d^{\,2}z_k
\prod\limits_{j=1}^{n+1}F(z_{2j-1},z_{2j})\ 
\frac{\Big(\sum_{l=1}^{2n+2}z_l\ -\ka\Big)\prod_{i>j}^{2n+2}(z_i-z_j)}
{\prod_{l=1}^{2n+2}(\ka-z_l)}
\nn\\
\fl&=&h_{n}\left\langle
\frac{\Tr J\ - \ka}{\det(\ka-J)}\right\rangle_{2n+2}\ .
\label{idt2proof}
\eea
Here 
we included the product into the Vandermonde determinant as before, as well
as the additional sum leading to the modified Vandermonde determinant
$\widetilde\Delta$ defined in the right-hand side of eq.\ (\ref{Vid}). In
the last step we 
simply need a slightly modified version of the identity
eq.\ (\ref{BidI}) which is derived in \ref{VIdII}.
The inclusion of the arbitrary constant
$c\neq0$ follows simply by shifting
$\ka\to\ka+c$ in the numerator but not in the denominator. This concludes
the derivation of all the integral representations of the SOP and their
Cauchy transforms. 
In principle the simple expectation values in eqs.~\eref{idq1} 
and \eref{idq2} as well as eqs.~(\ref{idt1}) and (\ref{idt2})
could be computed explicitly using 
supersymmetric {vectors depending on ordinary variables
  and Grassmannians}. In the explicit examples given in the
next section we shall give the resulting SOP only.

\sect{Examples for skew-orthogonal polynomials}\label{examples}

In this section we will give three examples of skew-orthogonal
polynomials in the complex plane: Hermite, Laguerre and Laguerre with
mass terms. Although the first of these were already known, our
derivation is new.  
The second and third are new examples.

In principle there are two different methods. In the first
of these we directly use the integral
representations; for the even polynomials these are the expectations
of a single 
determinant, eq.\ (\ref{idq1}), and for the odd polynomials the
expectations of a 
determinant multiplied by a trace,  eq.\ (\ref{idq2}). Both can 
{be calculated} in one
step by computing the expectation of the product of two determinants
(which is proportional 
to the kernel) and either taking limits or differentiating, and using
the fact that the 
determinant is the generating functional of all independent
{matrix invariants}.
The expectations can then be computed using Grassmannians, and because this
was already explicitly done in \cite{APSo} we can be very brief here. 

The second method follows the general setup outlined at the start of
section \ref{intrep}. 
Given the kernel in terms of general
polynomials eq.\ (\ref{Kdef}), and choosing them to be skew-orthogonal
eq.\ (\ref{skewdef}), the individual polynomials can be ``read off'' from
the kernel in eq.\ (\ref{kernel_sum_qq}) by differentiation (or taking limits):
\bea
q_{2n}(z)&=& h_n\frac{1}{(2n+1)!}\frac{\partial^{2n+1}}{\partial u^{2n+1}}
{\cal  K}_{2n+2}(u,z)\ =\ h_n\lim_{u\to\infty}
\frac{{\cal  K}_{2n+2}(u,z)}{u^{2n+1}}\nn\\
q_{2n+1}(z)&=& - h_n\frac{1}{(2n)!}\frac{\partial^{2n}}{\partial u^{2n}}
{\cal  K}_{2n+2}(u,z)\Big|_{u=0}\ +\ c\,q_{2n}(z)\ .
\label{qKdetermination}
\eea
This is possible whenever
the kernel has already been independently determined, e.g. by the above
procedure detailed in \cite{APSo} (see also \cite{Sommers2007} for
another method). In addition the norms $h_k$ can be read
off from the kernel as the leading coefficients. 

Of course both methods lead to the same answer. In the third example
the kernel including the masses as well as the partition function
itself have not previously been computed explicitly and
so this also constitutes a new result.

\subsection{Example I: skew-orthogonal Hermite polynomials}\label{sHermite}

In general the calculation of the expectation of a single
determinant (or the product of two determinants) is straightforward,
even without switching to an eigenvalue 
basis: we express the determinant as an
integral over anti-commuting (Grassmann) variables, and then the
Gaussian random 
matrices can be integrated out.  After a Hubbard-Stratonovich
transformation the anti-commuting variables can also be
integrated out. Because the procedure was carried out and
explained in detail for two determinants with $N_f=0$ in this model
in \cite{APSo} we only quote here the result for
our first example, the expectation with respect to model 
eq.\ (\ref{Zgin})\footnote{The double sum can be simplified by using the
  Christoffel-Darboux identity \cite{APSo}.}: 
\be
\fl \Big\langle \det(z-J)\,\det(u-J) \Big\rangle_N = N!\sum_{l=0}^N \tau^l
  \sum_{k=0}^l \frac{1}{k!\,2^k}\,H_k\left( \frac{z}{\sqrt{2\tau}}
  \right)\,H_k\left( \frac{u}{\sqrt{2\tau}} \right) 
\label{Hkernel}
\ee
where $\tau$ is the non-Hermiticity parameter, and the $H_k(z)$ are
the standard Hermite polynomials. 
Hence, for the even polynomials we can simply project out the second
determinant to give
\bea
\fl q_{2k}(z) & = & \lim_{u \rightarrow \infty} \frac{\Big\langle
  \det(z-J)\,\det(u-J) \Big\rangle_{2k}}{u^{2k}} 
\ =\ \left( \frac{\tau}{2} \right)^k\,H_{2k}\left(
\frac{z}{\sqrt{2\tau}} \right).
\label{Heven}
\eea
Here we used the following result to calculate the single term in the
double sum that survives the limiting process:
\be
\lim_{u \rightarrow \infty} 
\frac{1}{u^N}\,H_N\left( \frac{u}{\alpha} \right) \ = \ \left(
\frac{2}{\alpha} \right)^N.
\label{H_lim} 
\ee
This equation also implies that the even polynomials eq.\ (\ref{Heven})
are in monic normalisation as they should be, starting with $q_0(z)=1$. 
Alternatively we could of
course have differentiated eq.\ (\ref{Hkernel}) $N$ times with respect
to $u$.

For the odd polynomials, we use the fact that the determinant is the generating
functional for symmetric functions, and in particular for the trace:
\be 
\frac{1}{(N-1)!}\,\frac{\partial^{N-1}}{\partial u^{N-1}}
\det(u\eins_N-J) \Big|_{u=0} \ =\ - \Tr J\ ,  
\label{diff}
\ee
where $J$ is an $N\times N$ matrix, and $N\geq1$.
Applying this to eq.\ (\ref{Hkernel}) on the left-hand side allows us
to obtain  
$q_{2k+1}(z)$ from eq.\ (\ref{idq2}), for $k\geq1$:
\bea
\fl q_{2k+1}(z) & = & -\frac{1}{(2k-1)!}\,\frac{\partial^{2k-1}}{\partial
  u^{2k-1}}\Big\langle \det(z-J)\,\det(u-J) \Big\rangle_{2k} \Big|_{u=0} +
(z+c)q_{2k}(z) 
\nn \\ 
\fl & = & -(2k)\left( \sqrt{\frac{\tau}{2}} \right)^{2k-1}(\tau
+ 1)H_{2k-1}\left( \frac{z}{\sqrt{2\tau}} \right) \ +\ 
(z+c)\,\left(
\frac{\tau}{2} \right)^k H_{2k} \left( \frac{z}{\sqrt{2\tau}} \right)
\nn \\ 
\fl & = & \frac{\tau^{k+\frac12}}{2^{k+\frac12}}H_{2k+1}
\left( \frac{z}{\sqrt{2\tau}} \right) + 2k\, 
\frac{\tau^{k-\frac12}}{2^{k-\frac12}}H_{2k-1} \left(
\frac{z}{\sqrt{2\tau}} \right) + 
c\,\frac{\tau^k}{2^k}H_{2k} \left( \frac{z}{\sqrt{2\tau}}
\right) .
\label{Hodd}
\eea
In the first step only two terms survive the differentiation after
setting $u=0$; we used the following properties of Hermite
polynomials in addition to eq.\ (\ref{H_lim})
\bea 
\frac{d^{n-1}}{d z^{n-1}}\,H_n(z)&=& 2^n n!\,z  \label{H_diff} \\
H_{n+1}(z) & = & 2zH_n(z) - 2nH_{n-1}(z)\ ,\ \ 
\mbox{for}\ \ n\geq1\ ,  \label{H_rr}\
\eea
as well as the recurrence relation to simplify eq.\ (\ref{Hodd}) in the
last line. This form makes it more transparent that the arbitrary
addition of $cq_{2k}(z)$ is the only even Hermite polynomial appearing
in this example.
From eq.\ (\ref{H_lim}) it also follows that $q_{2k+1}(z)$
is in monic normalisation, and for $k=0$ we have $q_{1}(z)=z+c$ by definition. 
Defining 
\be
C_k(z)\equiv \left(\frac{\tau}{2}\right)^{\frac{k}{2}}H_{k}
\left( \frac{z}{\sqrt{2\tau}} \right)
\ee
we reobtain the following final simple result from \cite{Forrester07}
\bea
q_{2k}(z)   & = & C_{2k}(z),  \nn \\
q_{2k+1}(z) & = & C_{2k+1}(z) - 2kC_{2k-1}(z) - cC_{2k}(z)\ .
\label{Hfinal}
\eea

The norms $h_k=2(\tau+1)\sqrt{2\pi} \,(2k)!$
can be determined either by
direct calculation of the scalar product eq.\ (\ref{qdef}) 
of the SOP which we just obtained, as was done in \cite{Forrester07}, or by
computing the partition function\footnote{The lower order
  terms from combining eqs.\ (\ref{kernel_sum_qq}) and (\ref{idkernel})
will provide ratios of norms $h_N/h_k$ and thus the $k$-dependence only.}.

Let us emphasise that in our derivation the skew-orthogonality of the 
polynomials is automatically satisfied due to their
general integral representation eqs.\ (\ref{idq1}) and
(\ref{idq2}). This is in contrast to \cite{Forrester07}, 
where the skew-orthogonality was explicitly verified for the weights
eq.\ (\ref{wgin}) in the complex plane. 

\subsection{Example II: skew-orthogonal Laguerre polynomials}\label{sLaguerre}

In this subsection we turn to entirely new expressions for skew-orthogonal
Laguerre polynomials for our chiral model eq.\ (\ref{Zch}).
We will start with the so-called quenched case $(N_f=0)$ and then add
mass terms in the next subsection.

Expressed in terms of the $2n$ eigenvalues $z_j$ of $AB^T$ these are
the same as 
before; however, the matrix $\mcD$ in
eq.\ (\ref{Mdef}) also has $\nu$  generic zero eigenvalues, and so we
repeat the integral representations 
eqs.\ (\ref{idq1}) and (\ref{idq2}) for completeness, and also to make
contact with 
\cite{APSo}. For the even polynomials, we have 
\begin{equation}
\label{idq1ch}
\fl q_{2n}({z})=\frac{1}{z^{\nu/2}}\Big\langle
 \det(\sqrt{z}\eins_{4n+\nu}-\mcD)\Big\rangle_{4n+\nu}= \Big\langle
 \det({z}\eins_{2n}-AB^T)\Big\rangle_{2n}
\end{equation}
and for the odd polynomials
\begin{eqnarray}
\fl q_{2n+1}({z})&=&\frac{1}{z^{\nu/2}}
\left\langle 
\det(\sqrt{z}\eins_{4n+\nu}-\mcD)\left[\frac12\Tr \mcD^{\,2}+z+c\right]
\right\rangle_{4n+\nu}\nn\\
\fl&=& \Big\langle \det({z}\eins_{2n}-AB^T)\,\Tr AB^T\Big\rangle_{2n}
\ +\ (z+c)\, q_{2n}({z})
\,.\label{idq2ch}
\end{eqnarray}

The starting point for what we need for our calculations, 
namely the expectation of two determinants,
was again given in detail in \cite{APSo} and thus we merely state the result:
\bea
\fl&& 
\Big\langle \det({z}\eins_{2n}-AB^T)
 \det({u}\eins_{2n}-AB^T)\Big\rangle_{2n}
\nn\\
\fl &=&(2n)!\,(2n+\nu)!(4\mu^2\eta_+)^{4n}\sum_{l=0}^{2n} \left(
\frac{\eta_-}{\eta_+} \right)^{2l} 
\sum_{k=0}^{l}
\frac{k!}{(k+\nu)!}\,L_k^{\nu}\left( \frac{z}{4\mu^2\eta_-}
\right)L_k^{\nu}\left( \frac{u}{4\mu^2\eta_-} \right) ,
\label{Lkernel}
\eea
where we {recall the notation~\eref{mudef}}.

We thus obtain the even polynomials by simply projecting out the
second determinant
\bea
\fl q_{2k}(z) & = & \lim_{u \rightarrow \infty} 
\frac{1}{u^{2k}}
\Big\langle \det({z}\eins_{2k}-AB^T)
 \det({u}\eins_{2k}-AB^T)\Big\rangle_{2k}
\nn\\
\fl          & = & (4\mu^2\eta_-)^{2k}\,(2k)!\,L_{2k}^{\nu}\left(
\frac{z}{4\mu^2\eta_-} \right). 
\label{Leven}
\eea
Here we have used the following relation for the Laguerre polynomials
\be
\lim_{u \rightarrow \infty} \frac{1}{u^N}\,
L_N^{\nu}\left( \frac{u}{\al} \right) \ =\ \frac{(-1)^N}{N!\,\al^{N}}
\ ,
\label{L_lim}
\ee
which also confirms that the even polynomial is properly normalised.

For the odd polynomials we again need to take derivatives as in
eq.\ (\ref{diff}) to obtain for $k\geq1$
\bea
\fl q_{2k+1}(z) & = & -\frac{1}{(2k-1)!}\,\frac{\partial^{2k-1}}{\partial
  u^{2k-1}}
\Big\langle \det({z}\eins_{2k}-AB^T)
 \det({u}\eins_{2k}-AB^T)\Big\rangle_{2k} \, \Big|_{u=0}
\nn\\
\fl&&+(z+c)q_{2k}(z)
\nn \\ 
\fl & = & (4\mu^2\eta_-)^{2k+1}(2k)!\,(2k+\nu)
\left( 2kL_{2k}^{\nu}\left(\frac{z}{4\mu^2\eta_-} \right)
+\Big(1+\frac{\eta_+^2}{\eta_-^2} \Big)
L_{2k-1}^{\nu}\left(\frac{z}{4\mu^2\eta_-} \right)
\right)
\nn\\
\fl &&+(z+c)(4\mu^2\eta_-)^{2k}\,(2k)!\,
L_{2k}^{\nu}\left(\frac{z}{4\mu^2\eta_-} \right)
\nn\\
\fl&=& -(4\mu^2\eta_-)^{2k+1}\,(2k+1)!\,
L_{2k+1}^{\nu}\left(\frac{z}{4\mu^2\eta_-} \right)
+c'(4\mu^2\eta_-)^{2k}\,(2k)!
\,L_{2k}^{\nu}\left(\frac{z}{4\mu^2\eta_-} \right)
\nn\\
\fl&&
+(2k+\nu)(4\mu^2\eta_+)^2(4\mu^2\eta_-)^{2k-1}\,(2k)!
L_{2k-1}^{\nu}\left(\frac{z}{4\mu^2\eta_-} \right),
\label{Lodd}
\eea
where the new arbitrary constant
\be
c'\ \equiv\ c+(4\mu^2\eta_-)(4k^2+4k+1+(2k+1)\nu)
\ee
now depends on $k$, $\nu$ and $\mu$. 
The above result was obtained after using the corresponding relations for
Laguerre polynomials:
\bea
\fl
\frac{d^{n-1}}{d z^{n-1}}L_n^{\nu}(z) & = & (-1)^n(z-(n+\nu))  
\label{L_diff}\\
\fl (n+1)L_{n+1}^{\nu}(z) & = & (2n+\nu+1-z)L_n^{\nu}(z) -(n+\nu)
L_{n-1}^{\nu}(z)\ ,\mbox{for}\ \ n\geq1\ .
\label{L_rr} 
\eea
It is easy to see that the polynomials are again monic, due to
eq.\ (\ref{L_lim}). This once more fixes $q_1(z)=z+c$.
We can now define
\bea
C_k^{\nu}(z) & \equiv & (4\mu^2\eta_-)^{k}\,k!\,
L_k^{\nu}\left( \frac{z}{4\mu^2\eta_-} \right)
\eea
allowing us to write 
\bea
\fl q_{2k}(z)    & = & +C_{2k}^{\nu}(z),  \nn \\
\fl q_{2k+1}(z)  & = & -C_{2k+1}^{\nu}(z) +
(1+\mu^2)^2\,(2k)(2k+\nu)C_{2k-1}^{\nu}(z) + c'C_{2k}^{\nu}(z)\ , 
\label{Lfinal}
\eea 
giving the new skew-orthogonal Laguerre polynomials up to an arbitrary
constant. 
The final result compares with the similar form of eq.\ (\ref{Hfinal}).

For the norms we find 
$h_k= 8\pi(4\mu^2)
(2k)!\,(2k+\nu)!\,(4\mu^2\eta_+)^{4k+\nu+1}$ where the $k$-dependence
again follows from the ratio of the norms $h_N/h_k$,
see eqs.\ (\ref{idkernel}) and (\ref{kernel_sum_qq}), whereas the
overall constant factor can be deduced from the partition function,
see eq.\ (3.46) in \cite{APSoII}, and taking the ratio for consecutive
values of $N$. 

\subsection{Example III: inclusion of mass terms in the 
chiral model}\label{smass}

Our third example gives the SOP again for weights including 
$N_f > 0$ mass terms, which is also called the unquenched case. We will 
exemplify this using the chiral model eq.\ (\ref{Zch}) where such terms are
more common due to applications in QCD. However, the same insertion
of mass terms can be done in the non-chiral model eq.\ (\ref{Zgin})
following the same lines. 

Our main point here will be to express the SOP for $N_f > 0$ in
terms of the SOP for $N_f=0$ (and the corresponding kernel), which we
have already calculated. To indicate which polynomials we are referring to
we will use a superscript, as in $q_{k}^{(N_f)}(z)$, and
correspondingly for the kernel and expectations. 

Our derivation relies heavily on \cite{KGII} where all the expectation
values of products and ratios of characteristic polynomials (or determinants)
have been expressed in terms of Pfaffian expressions of matrices
containing a small number of building blocks; in our case these
building blocks will be the quenched ($N_f=0$) SOP and kernel.

To begin we first express the unquenched integral representations
eqs.\ (\ref{idq1}) 
and (\ref{idq2}) in terms of ratios of quenched expectations.
For the even polynomials we have 
\bea
\label{idq1chNf}
\fl q_{2n}^{(N_f)}({z})&=& \Big\langle
 \det({z}\eins_{2n}-AB^T)\Big\rangle_{2n}^{(N_f)}
\\
\fl &=& \frac{\Big\langle
 \det({z}\eins_{2n}-AB^T)\prod_{f=1}^{N_f}\det({m^2_f}\eins_{2n}-AB^T)
\Big\rangle_{2n}^{(0)}}{
\Big\langle \prod_{f=1}^{N_f}\det({m^2_f}\eins_{2n}-AB^T)
\Big\rangle_{2n}^{(0)}}\ ,
\nn
\eea
and similarly for the odd polynomials we have 
\bea
\fl q_{2n+1}^{(N_f)}({z})
 &=& \frac{\Big\langle
 \det({z}\eins_{2n}-AB^T) \, \Tr AB^T
\prod_{f=1}^{N_f}\det({m^2_f}\eins_{2n}-AB^T)
\Big\rangle_{2n}^{(0)}}{
\Big\langle \prod_{f=1}^{N_f}\det({m^2_f}\eins_{2n}-AB^T)
\Big\rangle_{2n}^{(0)}}\ +(z+c)q_{2n}^{(N_f)}({z}).
\nn\\
\fl&&
\label{idq2chNf}
\eea
We will now give all the building blocks for these expressions.
The first building block in the denominator, 
the expectation value of the mass term, simultaneously provides us
with the massive partition function itself:
\be
\frac{{\cal Z}^{(N_f)}_{ch\, 2N}(\{m\})}{{\cal Z}^{(0)}_{ch\, 2N}}
=\prod_{f=1}^{N_f}m_f^\nu\ 
\Big\langle \prod_{f=1}^{N_f}\det({m^2_f}\eins_{2N}-AB^T)
\Big\rangle_{2N}^{(0)}\ .
\label{Zmass}
\ee
The masses to the power of $\nu$, the number of generic zero
eigenvalues, of course cancel in the ratios for the $q_k^{(N_f)}(z)$ above. 
Using the results from \cite{KGII} and expressing the expectation
values there in terms of our quenched kernel and even SOP we obtain:
\bea
\fl\frac{{\cal Z}^{(N_f)}_{ch\, 2N}(\{m\})}{{\cal Z}^{(0)}_{ch\, 2N}}
&=&\frac{\prod_{f=1}^{N_f}m_f^\nu}{\Delta_{N_f}(\{m^2\})}
(-)^{N_f/2}\prod_{j=N}^{N+(N_f-2)/2}h_j^{(0)} 
\underset{1\leq f,g\leq N_f}{\Pf}
\left[{\cal K}^{(0)}_{2N+N_f}(m_f^2,m_g^2)\right]
\nn\\
\fl&&
\ \ \ \ \ \ \ \ \ \ \ \ \ \ \ \ \ \ \ \ \ \ \ \ \ \ \ \ \ 
\ \ \ \ \ \ \ \ \ \ \ \ \ \ \ \ \ \ \ \ \ \ \ \ \ \ \ \ \ 
\ \ \ \ \ \ \ \ \ \ \ \ \ 
N_f\ \mbox{even}
\nn\\
\fl\frac{{\cal Z}^{(N_f)}_{ch\, 2N}(\{m\})}{{\cal Z}^{(0)}_{ch\, 2N}}
&=&\frac{\prod_{f=1}^{N_f}m_f^\nu}{\Delta_{N_f}(\{m^2\})}
(-)^{(N_f-1)/2}\prod_{j=N}^{N+(N_f-3)/2}h_j^{(0)}
\nn\\
\fl&&\times
\underset{1\leq f,g\leq N_f}{\Pf}\left[
\begin{array}{cc}
0 & q_{2N+N_f-1}^{(0)}(m_g^2)\\
-q_{2N+N_f-1}^{(0)}(m_f^2)  & {\cal K
}^{(0)}_{2N+N_f-1}(m_f^2,m_g^2)\\ 
\end{array}
\right]
{,\ \ N_f\ \mbox{odd}}
\label{ZNF}
\eea
where we have to distinguish even and odd numbers of flavours $N_f$.
The product over the norms in the prefactor is equal to unity when the
upper limit is $N-1$. Compared to \cite{KGII} we have used the
following identity
\be
\fl
\underset{1\leq f,g\leq N_f}{\Pf}\!\left[\!
\begin{array}{cc}
0 & q_{2M}^{(0)}(m_g^2)\\
-q_{2M}^{(0)}(m_f^2)  & {\cal K
}^{(0)}_{2M+2}(m_f^2,m_g^2)\\ 
\end{array}
\right]
=
\underset{1\leq f,g\leq N_f}{\Pf}\!\left[\!
\begin{array}{cc}
0 & q_{2M}^{(0)}(m_g^2)\\
-q_{2M}^{(0)}(m_f^2)  & {\cal K
}^{(0)}_{2M}(m_f^2,m_g^2)\\ 
\end{array}
\right]
\label{Pfid}
\ee
which can be easily seen by adding multiples of the first row and
column to the remaining rows and columns, in order to eliminate the
leading SOP in the kernels and hence shifting their index down by two.
This result for the partition function (or expectation values of
characteristic polynomials) precisely equals the corresponding result
for $\beta=4$ in \cite{AB}\footnote{Notice a typo in \cite{AB} in eq. (2.8) 
  compared to the correct Theorem 1 in eq. (3.1) there.}.

The even polynomials now easily follow from eq. (\ref{ZNF}), by choosing one
of the masses to be the argument. We obtain 
\be
\fl q_{2N}^{(N_f)}({z})=
\frac{\Pf
\left[
\begin{array}{ccc}
0                   & q_{2N+N_f}^{(0)}(z)&q_{2N+N_f}^{(0)}(m_g^2)\\
-q_{2N+N_f}^{(0)}(z)& 0                 & {\cal K }^{(0)}_{2N+N_f}(z,m_g^2)\\
-q_{2N+N_f}^{(0)}(m_f^2) & {\cal K }^{(0)}_{2N+N_f}(m_f^2,z) 
& {\cal K }^{(0)}_{2N+N_f}(m_f^2,m_g^2)\\
\end{array}
\right]
}{\prod_{f=1}^{N_f}(z-m_f^2)\ \Pf
\left[{\cal K
    }^{(0)}_{2N+N_f}(m_f^2,m_g^2)\right]}
\label{qNFe}
\ee
{for $N_f$ even}. Here and in the following we suppress the
indices of the Pfaffian which run from $1$ to $N_f$ in both the even
and odd cases. For $N_f$ odd we obtain
\be
\fl q_{2N}^{(N_f)}({z})=-\ 
\frac{h^{(0)}_{N+(N_f-1)/2}
\Pf
\left[
\begin{array}{cc}
0                                 & {\cal K }^{(0)}_{2N+N_f+1}(z,m_g^2)\\
{\cal K }^{(0)}_{2N+N_f+1}(m_f^2,z) & {\cal K }^{(0)}_{2N+N_f+1}(m_f^2,m_g^2)\\
\end{array}
\right]
}{\prod_{f=1}^{N_f}(z-m_f^2)\ \Pf
\left[
\begin{array}{cc}
0 & q_{2N+N_f-1}^{(0)}(m_g^2)\\
-q_{2N+N_f-1}^{(0)}(m_f^2)  & {\cal K }^{(0)}_{2N+N_f-1}(m_f^2,m_g^2)\\
\end{array}
\right]}.
\label{qNFo}
\ee

Next we determine the massive kernel using eq. (\ref{idkernel})
\bea
\fl&& 
{{\cal K }^{(N_f)}_{2N} (z,u)}
\ =\ \frac{(z-u)}{h_{N-1}^{(N_f)}}
\Big\langle \det({z}\eins_{2N}-AB^T)\det({u}\eins_{2N}-AB^T)
\Big\rangle^{(N_f)}_{2N-2}
\label{kernelNF}\\
\fl&&= \frac{(z-u)}{h_{N-1}^{(N_f)}}
\frac{\Big\langle
 \det({z}\eins_{2n}-AB^T)\det({u}\eins_{2N}-AB^T)
\prod_{f=1}^{N_f}\det({m^2_f}\eins_{2n}-AB^T)
\Big\rangle_{2N-2}^{(0)}}{
\Big\langle \prod_{f=1}^{N_f}\det({m^2_f}\eins_{2n}-AB^T)
\Big\rangle_{2N-2}^{(0)}}\ .
\nn
\eea
Using eq.\ (\ref{ZNF}) with two extra masses we obtain 
for $N_f$ even 
\bea
\fl 
{{\cal K }^{(N_f)}_{2N} (z,u)}
&=&
-\ \frac{\Pf
\left[
\begin{array}{ccc}
0 & {\cal K }^{(0)}_{2N+N_f}(u,z) & {\cal K }^{(0)}_{2N+N_f}(u,m_g^2)
\\
{\cal K }^{(0)}_{2N+N_f}(z,u) & 0  & {\cal K }^{(0)}_{2N+N_f}(z,m_g^2)
\\
{\cal K }^{(0)}_{2N+N_f}(m_f^2,u) & {\cal K }^{(0)}_{2N+N_f}(m_f^2,z) 
& {\cal K }^{(0)}_{2N+N_f}(m_f^2,m_g^2)\\
\end{array}
\right]
}{\prod_{f=1}^{N_f}(z-m_f^2)(u-m_f^2) \ 
\Pf
\left[{\cal K
    }^{(0)}_{2N+N_f}(m_f^2,m_g^2)\right]}\ .
\label{KNFe}
\eea
Here the mass dependent inverse norm $1/h_{N-1}^{(N_f)}$ has been
  eliminated using the following identity, leading to a shift in the
  index of the kernels in the denominator by $+2$. Following eq. (\ref{Zprod})
we can write
\bea
\fl\frac{h_{N-1}^{(N_f)}}{h_{N-1}^{(0)}}&=&
\frac{{\cal Z}^{(N_f)}_{ch\, 2N}(\{m\})}{{\cal Z}^{(0)}_{ch\, 2N}}
\ \frac{{\cal Z}^{(0)}_{ch\, 2N-2}}{{\cal Z}^{(N_f)}_{ch\,
    2N-2}(\{m\})}
\label{hratio}\\
\fl&=&
\left\{
\begin{array}{ll}
\displaystyle\frac{h^{(0)}_{N+(N_f-2)/2}}{h^{(0)}_{N-1}}
 \frac{{\Pf}
\left[{\cal K}^{(0)}_{2N+N_f}(m_f^2,m_g^2)\right]}{{\Pf}
\left[{\cal K}^{(0)}_{2N-2+N_f}(m_f^2,m_g^2)\right]}&,\  N_f\ \mbox{even}\\
&\\
\displaystyle\frac{h^{(0)}_{N+(N_f-3)/2}}{h^{(0)}_{N-1}}
\frac{{\Pf}\left[
\begin{array}{cc}
0 & q_{2N+N_f-1}^{(0)}(m_g^2)\\
-q_{2N+N_f-1}^{(0)}(m_f^2)  & {\cal K
}^{(0)}_{2N+N_f-1}(m_f^2,m_g^2)\\ 
\end{array}
\right]}{{\Pf}\left[
\begin{array}{cc}
0 & q_{2N+N_f-3}^{(0)}(m_g^2)\\
-q_{2N+N_f-3}^{(0)}(m_f^2)  & {\cal K
}^{(0)}_{2N+N_f-3}(m_f^2,m_g^2)\\ 
\end{array}
\right]}&,\  N_f\ \mbox{odd}.\\
\end{array}
\right.\nn
\eea
Likewise for $N_f$ odd we have
\bea
\fl 
&&{{\cal K }^{(N_f)}_{2N} (z,u)}
=
\frac{-1}
{\prod_{f=1}^{N_f}(z-m_f^2)(u-m_f^2)
\Pf
\left[
\begin{array}{cc}
0 & q_{2N+N_f-1}^{(0)}(m_g^2)\\
-q_{2N+N_f-1}^{(0)}(m_f^2)  & {\cal K }^{(0)}_{2N+N_f-1}(m_f^2,m_g^2)\\
\end{array}
\right]
}
\nn\\
\fl&& 
\times
\Pf
\left[
\begin{array}{cccc}
0&q_{2N+N_f-1}^{(0)}(z) & q_{2N+N_f-1}^{(0)}(u)&q_{2N+N_f-1}^{(0)}(m_g^2)\\
-q_{2N+N_f-1}^{(0)}(z) &0 & {\cal K }^{(0)}_{2N+N_f-1}({z,u}) 
&{\cal K }^{(0)}_{2N+N_f-1}({z},m_g^2)
\\
-q_{2N+N_f-1}^{(0)}(u)&{\cal K }^{(0)}_{2N+N_f-1}({u,z}) & 0 
& {\cal K }^{(0)}_{2N+N_f-1}({u},m_g^2)
\\
-q_{2N+N_f-1}^{(0)}(m_f^2)&{\cal K }^{(0)}_{2N+N_f-1}(m_f^2,{z}) 
& {\cal K }^{(0)}_{2N+N_f-1}(m_f^2,{u}) 
& {\cal K }^{(0)}_{2N+N_f-1}(m_f^2,m_g^2)
\\
\end{array}
\right].\nonumber\\
\fl&&\label{KNFo}
\eea
Let us pause with a few remarks. Following \cite{APSoII} 
these expressions for the massive
kernel determine all massive eigenvalue
correlation functions for $2N$, in terms of the known quenched kernel and the
quenched even SOP that were given in Example II above.
In particular it is transparent that even for finite $N$ the
unquenched kernel (when properly normalised by the massive weight) is
given by the quenched kernel plus some correction terms. The same
structure thus prevails for all unquenched eigenvalue correlation functions.

For $2N+1$, correction terms to the massive kernel will also include the
massive SOP $q_{2N}^{(N_f)}(z)$, when following
e.g. \cite{ForresterMays}. Thus all massive eigenvalue
correlation functions follow in this case as well.

As the final step we will give the massive odd  SOP
$q_{2N+1}^{(N_f)}(z)$. Here we will follow
eq.\ (\ref{qKdetermination}) and determine them from the kernel, 
rather than eq.\ (\ref{diff}).
As an aside, above we could have alternatively determined the kernel first 
and then the even SOP from eq.\ (\ref{qKdetermination}) as well.
A slight generalisation of eq.\ (\ref{qKdetermination}) reads
\be
\fl q_{2n+1}^{(N_f)}(z)= - h_n^{(N_f)}\frac{1}{(2n+k)!}
\frac{\partial^{2n+k}}{\partial  u^{2n+k}}
\left.\left(\prod_{l=1}^k(u-a_l)
{\cal  K}_{2n+2}^{(N_f)}(u,z)\right)\right|_{u=0}\ +\ c'\,q_{2n}^{(N_f)}(z)
\label{qKdet2}
\ee
where $k \ge 0$, and the $a_l$ are some arbitrary constants. It is
easy to see that  
we only get a non-vanishing result when $2n$ or $2n+1$ of the
derivatives act on the kernel and not the prefactor. {This is true
  because the function in the brackets is a polynomial of order
  $2n+k+1$ in the variable $u$. Hence, the differentiation yields the
  coefficients of the monomials of order $2n$ and $2n+1$ in the
  variable $u$ of the kernel ${\cal  K}_{2n+2}^{(N_f)}$.} 

In choosing $k=N_f$ and the $a_l=m_l^2$ we can use this relation to
cancel  the factor $\prod_{f=1}^{N_f}(u-m_f^2)$ in the denominator of
eq.\ (\ref{KNFe}) 
that would otherwise have to be differentiated as well. 
We thus obtain from eq.\ (\ref{qKdet2}) and (\ref{KNFe}) that
\bea
\fl q_{2N+1}^{(N_f)}(z)&=& - h_N^{(N_f)}\frac{1}{(2N+N_f)!}
\frac{\partial^{2N+N_f}}{\partial  u^{2N+N_f}}
\left.\left(\prod_{f=1}^{N_f}(u-m_f^2)
{\cal  K}_{2N+2}^{(N_f)}(u,z)\right)\right|_{u=0}\ +\ c'\,q_{2N}^{(N_f)}(z)
\nn\\
\fl &=&
\frac{\Pf
\left[
\begin{array}{ccc}
0 & q^{(0)}_{2N+N_f+1}(z) & q^{(0)}_{2N+N_f+1}(m_g^2)
\\
-q^{(0)}_{2N+N_f+1}(z)& 0  & {\cal K }^{(0)}_{2N+N_f}(z,m_g^2)
\\
-q^{(0)}_{2N+N_f+1}(m_f^2)& {\cal K }^{(0)}_{2N+N_f}(m_f^2,z) 
& {\cal K }^{(0)}_{2N+N_f}(m_f^2,m_g^2)\\
\end{array}
\right]
}{\prod_{f=1}^{N_f}(z-m_f^2)\ 
\Pf
\left[{\cal K}^{(0)}_{2N+N_f}(m_f^2,m_g^2)\right]}
+c'q_{2N}^{(N_f)}(z)
\nn\\
\fl&& 
\ \ 
\mbox{for}\ \ N_f\ \mbox{even}.
\eea
Here we have pulled the derivatives inside the Pfaffian leading to the
quenched SOP of shifted odd index, and used eq.\ (\ref{hratio})
and the identity corresponding to eq.\ (\ref{Pfid}) for the odd polynomials.
For odd $N_f$ we obtain
\bea
\fl &&q_{2N+1}^{(N_f)}(z)=
\frac{1
}{\prod_{f=1}^{N_f}(z-m_f^2)\ 
\Pf
\left[
\begin{array}{cc}
0 & q_{2N+N_f-1}^{(0)}(m_g^2)\\
-q_{2N+N_f-1}^{(0)}(m_f^2)  & {\cal K }^{(0)}_{2N+N_f-1}(m_f^2,m_g^2)\\
\end{array}
\right]}
\nn\\
\fl&& 
\times
\Pf
\left[
\begin{array}{cccc}
0&q_{2N+N_f+1}^{(0)}(z) & \tilde{c} &q_{2N+N_f+1}^{(0)}(m_g^2)\\
-q_{2N+N_f+1}^{(0)}(z) &0 & q^{(0)}_{2N+N_f-1}({z}) 
&{\cal K }^{(0)}_{2N+N_f+1}({z},m_g^2)
\\
-\tilde{c}&-q^{(0)}_{2N+N_f-1}({z}) & 0 
& q^{(0)}_{2N+N_f-1}(m_g^2)
\\
-q_{2N+N_f+1}^{(0)}(m_f^2)&{\cal K }^{(0)}_{2N+N_f+1}(m_f^2,{z}) 
& -q^{(0)}_{2N+N_f-1}(m_f^2) 
& {\cal K }^{(0)}_{2N+N_f+1}(m_f^2,m_g^2)
\\
\end{array}
\right]\nn\\
\fl&&
+c'q_{2N}^{(N_f)}(z)\ \ 
\mbox{for}\ \ N_f\ \mbox{odd}.
\label{qoNFo}
\eea
The constants $\tilde{c}$ in the Pfaffian in the numerator can be
absorbed into the even polynomial $c'q_{2N}^{(N_f)}(z)$, which can be
seen as follows. Just as the determinants of two matrices that only
differ by a single row (or column) can be added, a similar statement
holds for Pfaffians: due to linearity the Pfaffians of two
anti-symmetric matrices that only differ by a single row and its
transposed column can be added. We can thus split off the
$\tilde{c}$-part from the Pfaffian above to obtain
\bea
&&\Pf
\left[
\begin{array}{cccc}
0&q_{2N+N_f+1}^{(0)}(z) & \tilde{c} &q_{2N+N_f+1}^{(0)}(m_g^2)\\
-q_{2N+N_f+1}^{(0)}(z) &0 & 0
&{\cal K }^{(0)}_{2N+N_f+1}({z},m_g^2)
\\
-\tilde{c}&0 & 0 & 0
\\
-q_{2N+N_f+1}^{(0)}(m_f^2)&{\cal K }^{(0)}_{2N+N_f+1}(m_f^2,{z}) 
& 0
& {\cal K }^{(0)}_{2N+N_f+1}(m_f^2,m_g^2)
\\
\end{array}
\right]\nn\\
\fl&&=-\ 
\Pf
\left[
\begin{array}{cccc}
0& \tilde{c} &0&0\\
-\tilde{c}&0 & 0 & 0\\
0&0 & 0
&{\cal K }^{(0)}_{2N+N_f+1}({z},m_g^2)
\\
0&0
&{\cal K }^{(0)}_{2N+N_f+1}(m_f^2,{z}) 
& {\cal K }^{(0)}_{2N+N_f+1}(m_f^2,m_g^2)
\\
\end{array}
\right]
\eea
which is proportional to the numerator of the even polynomials with
odd $N_f$ in eq.\ (\ref{qNFo}). Thus the final result for the odd
polynomial with odd $N_f$ is eq.\ (\ref{qoNFo}) with $\tilde{c}=0$ and
$c'\to c^{''}$.
This ends our third example for the SOP and
kernel including masses.

Similar expressions could be given for the non-chiral model
eq.\ (\ref{Zgin}), as well as for the 
Cauchy transforms of the unquenched SOP.


\sect{Conclusions}

In this paper we have completed the analysis of the set of (skew-)
orthogonal polynomials  
in the complex plane 
that apply to the chiral extensions of the three elliptic Ginibre
ensembles. By constructing an explicit integral representation we
found a new set of skew-orthogonal Laguerre polynomials in the complex
plane which provide an alternative method of 
solving the chiral ensemble with
real asymmetric elements. Our integral representation is also valid
for the real elliptic Ginibre ensemble; in fact, it holds for arbitrary
weight functions $g$ and $h$ in these two classes. 
Furthermore we also gave a new integral representation of
the Cauchy transforms of these polynomials which holds not only for the
two symmetry classes with real matrix elements ($\beta=1$) 
but also for quaternion real matrix elements ($\beta=4$).

An important ingredient for our results was a proof that the
probability distribution in the partition function factorises for
$\beta=1$. 
This offers another unifying view of the non-Hermitian $\beta=1$ and
$\beta=4$ symmetry 
classes, both chiral and non-chiral.

There are many more non-Hermitian ensembles, in addition to the three
Ginibre classes and their chiral counterparts, all six of which have  
now been solved.
It thus remains an open question whether corresponding sets of
orthogonal or skew-orthogonal 
polynomials exist for the other ensembles. It is possible that,
just as in the real case, the known polynomials also apply to some of
these other
non-Hermitian symmetry classes, once a complex eigenvalue
representation has been found for them.  

As an application of our results we have shown how to 
construct the skew-orthogonal polynomials and the kernel
when including $N_f$ characteristic polynomials or mass terms into our
models. These constitute the building blocks for the massive partition
function and eigenvalue density correlation functions. 
Consequently this will
allow us to study the complex Dirac operator spectrum for Quantum
Chromodynamics with two colours and non-vanishing quark chemical potential,
both in the low and high density phases. The study of the large-$N$
limit needed for this partly follows from the known quenched
$N_f=0$ results and is
left for future investigations.


\ \\[1ex] 
\textbf{Acknowledgements:} We thank the Niels Bohr
Foundation for financial support 
(G.A.), as well as the Niels Bohr Institute and International
Academy for the warm hospitality (G.A., M.J.P.). 
We also acknowledge support by the Deutsche Forschungsgemeinschaft within
  Sonderforschungsbereich Transregio 12 ``Symmetries and Universality
  in Mesoscopic Systems'' (M.K.) and support by an EPSRC doctoral
  training grant  (M.J.P.). Furthermore we would like to thank Tilo
  Wettig and Takuya Kanazawa for useful exchanges.

\begin{appendix}
\sect{Generalisation of the de Bruijn integral formula}\label{GendB}

In this appendix we slightly generalise the standard de Bruijn
integral formula that reads
\bea
\fl&&\prod\limits_{j=1}^{n} \int\limits_{\mathbb{C}} d^{\,2}z_j \, w(z_j)
\det_{1\leq a\leq 2n;\,1\leq l\leq n}
[\{f_a(z_{l}),g_a(z_{l})\}]\nn\\
\fl=&&{n!}\ \underset{1\leq a,b\leq 2n}{\Pf}\left[
\int\limits_{\mathbb{C}}d^{\,2}u\, w(u)[f_a(u)g_b(u)-f_b(u)g_a(u)]
\right]\ .
\eea
Here $w(z)$ is a weight function in the complex plane and $f$ and $g$ are
functions such that the integrals exist. The proof is usually done by a
Laplace expansion into $2\times 2$ blocks that each depend on a single
variable $z_l$.

If we start out with $2n$ instead of $n$ integrations over a product
of an anti-symmetric weight $F(u,v)$ 
and let $f$ and $g$ depend on different variables, we
have on the left-hand side 
\bea
\fl&&\prod_{k=1}^{2n} \int\limits_{\mathbb{C}} d^{\,2}z_k
\prod\limits_{j=1}^{n}F(z_{2j-1},z_{2j})
\det_{1\leq a\leq 2n;\,1\leq b\leq n}
[\{f_a(z_{2b-1}),g_a(z_{2b})\}]\nn\\
\fl&=&
\prod\limits_{j=1}^{n}\int\limits_{\mathbb{C}^2}d^{\,2}z_{2j-1}d^{\,2}z_{2j}
F(z_{2j-1},z_{2j})
\sum_{\sigma}(-)^\sigma
\prod_{j=1}^n
\det\left[
\begin{array}{cc}
f_{\sigma(2j-1)}(z_{2j-1})&g_{\sigma(2j-1)}(z_{2j})\\
f_{\sigma(2j)}(z_{2j-1})&g_{\sigma(2j)}(z_{2j})\\ 
\end{array}\right]
\nn\\
\fl&=&n!\ \underset{1\leq a,b\leq 2n}{\Pf}\left[ \,\, 
\int\limits_{\mathbb{C}^2}
d^{\,2}u \, d^{\,2}vF(u,v)[f_a(u)g_b(v)-f_b(u)g_a(v)]
\right]\ .
\eea
Here $(-)^\sigma$ is the sign of the permutation of the $2n$ variables,
and the sum is over all $(2n)!$ permutations which satisfy the
restriction $\sigma(1)<\sigma(2)<\ldots<\sigma(2n)$. Note that each
pair $\{z_{2j-1},z_{2j}\}$ only appears in one subdeterminant. This
gives the Pfaffian as a result (see e.g. \cite{Mehta} for a definition).

\sect{Relation to a modified Vandermonde determinant}\label{VIdI}

In this appendix we prove an identity related to Vandermonde determinants,
which is needed to derive the integral representation eq.\ (\ref{idq2}) for the
odd polynomials $q_{2n+1}(z)$. For completeness 
we repeat the relation eq.\ (\ref{Vid}) which is to be shown here,
\be
\sum_{i=1}^{N}z_i\ \Delta_{N}(\{z\})\ =\ \det\left[
\begin{array}{lll}
1  &\ldots&1\\
z_1&\ldots&z_{N}\\
\vdots&&\vdots\\
z_1^{N-2}&\ldots&z_{N}^{N-2}\\
z_1^{N}&\ldots&z_{N}^{N}\\
\end{array}
\right]\equiv \widetilde{\Delta}_{N}(\{z\})\ .
\label{VidA}
\ee
Using the second representation from eq.\ (\ref{Vandermonde}),
$\Delta_{N}(\{z\})=\det_{1\leq a,b\leq N}[z_a^{b-1}],$
one can see that the modified Vandermonde determinant
in eq.\ (\ref{VidA})
has a mismatch of 1 in the powers in the last row compared to the 
Vandermonde determinant.

Our proof uses that $\widetilde{\Delta}_{N}(\{z\})$ is simply the
coefficient of power $u^{N-1}$ in a Leibniz expansion of the
Vandermonde determinant ${\Delta}_{N+1}(\{z\},u)$ of size $N+1$ 
with respect to the last column in 
the extra variable $u$. This term can be singled out by a
differentiation, 
\be
\frac{1}{(N-1)!}\left.\frac{\partial^{N-1}}{\partial u^{N-1}}\right|_{u=0}
\Delta_{N+1}(\{z\},u)=-\widetilde{\Delta}_{N}(\{z\})\ .
\label{2}
\ee
On the other hand, using again eq.\ (\ref{Vandermonde}) that 
$\Delta_{N}(\{z\})=\prod_{j>k}^N(z_j-z_k)$,
one can write
\begin{equation}
 \Delta_{N+1}(\{z\},u)=\prod\limits_{a=1}^{N}(u-z_a)
\Delta_{N}(\{z\}).\label{3}
\end{equation}
Combining eqs.~\eref{2} and \eref{3} we obtain the result~\eref{VidA}.

\sect{Cauchy-type identity for the modified Vandermonde
  determinant}\label{VIdII}

In this appendix we prove the following identity, 
\be
\fl \det_{1\leq a\leq 2n+2;\,1\leq b\leq 2n}\left[
\{z_a^{b-1}\}\Big|z_a^{2n+1}\Big|\frac{1}{\ka-z_a}\right]
\ =\ 
\frac{\Big(\sum_{k=1}^{2n+2}z_k\ -\ka\Big)\prod_{i>j}^{2n+2}(z_i-z_j)}
{\prod_{l=1}^{2n+2}(\ka-z_l)}\ ,
\label{BidII}
\ee
which is a slight modification of identity (\ref{BidI}) with a
mismatch by one power in the last but one column. In fact we will use
the identity (\ref{BidI}) to prove the above. Expanding the left-hand
side with respect to the last but one column we have 
\bea
\fl 
&&\sum_{k=1}^{2n+2}(-)^{2n+2-k-1}z_{k}^{2n+1}
\det_{1\leq a\neq k\leq 2n+2;\,1\leq b\leq 2n}\left[
\{z_a^{b-1}\}\Big|\frac{1}{\ka-z_a}\right]
\nn\\
\fl &=&(-)\sum_{k=1}^{2n+2}(-)^{2n+2-k}z_{k}^{2n+1}
\frac{\prod_{i>j;\, i,j\neq k}^{2n+2}(z_i-z_j)}
{\prod_{l\neq k}^{2n+2}(\ka-z_l)}\ \frac{(\ka-z_k)}{(\ka-z_k)}
\nn\\
\fl &=&(-)\frac{\ka{\Delta}_{2n+2}(\{z\})-\widetilde{\Delta}_{2n+2}(\{z\})}
{\prod_{l=1}^{2n+2}(\ka-z_l)}\nn\\
\fl 
&=&\frac{\Big(\sum_{k=1}^{2n+2}z_k\ -\ka\Big)\prod_{i>j}^{2n+2}(z_i-z_j)}
{\prod_{l=1}^{2n+2}(\ka-z_l)}\ .
\eea
In the second step we used the identity (\ref{BidI}) and multiplied
by unity to complete the product in the denominator. For the
numerator we obtain the modified Vandermonde determinant 
$\widetilde{\Delta}$ from 
eq.\ (\ref{Vid}) and a
proper Vandermonde determinant, both of size $2n+2$, 
after resumming the expansion. Writing out
explicitly $\widetilde{\Delta}_{2n+2}$ from the left-hand side of eq.\
(\ref{VidA}) 
yields the right-hand side of our identity (\ref{BidII}).

\end{appendix}




\section*{References}
\begin{thebibliography}{99}

\bibitem{Mehta}
M.L. Mehta, {\it Random Matrices}, Academic Press, Third
Edition, London (2004).

\bibitem{Ghosh} S. Ghosh, {\it Skew-Orthogonal Polynomials and Random
  Matrix Theory}, CRM Monographs Series, vol. {\bf 28}, 
and AMS publication (2009).

\bibitem{Taro} T. Nagao, J. Stat. Phys. {\bf 129} (2007) 1137 
[arXiv:0708.2036].

\bibitem{Deift} P. Deift and D. Gioev, {\it Random Matrix Theory:
  Invariant Ensembles and Universality}, 
Courant Lecture Notes, vol. {\bf 18} and AMS
  publication (2009).



\bibitem{PDF} P. Di Francesco, M. Gaudin, C. Itzykson and F. Lesage,
Int. J. Mod. Phys. {\bf A9} (1994) 4257 [arXiv:hep-th/9401163].

\bibitem{Kan02}
E. Kanzieper, 
J. Phys. {\bf A
35} (2002) 6631
[arXiv:cond-mat/0109287].

\bibitem{Forrester07} P.J. Forrester and T. Nagao,
  Phys. Rev. Lett. {\bf 99} (2007) 050603 [arXiv:0706.2020]; 
\newline
J. Phys. {\bf A41} (2008) 375003 [arXiv:0806.0055].

\bibitem{James} J.C. Osborn,
Phys. Rev. Lett. {\bf 93} (2004) 222001
[arXiv:hep-th/0403131].

\bibitem{A05}
G. Akemann,
  Nucl.\ Phys. {\bf B730} (2005) 253    [arXiv:hep-th/0507156].

\bibitem{APSo}
G. Akemann, M.J. Phillips and H.-J. Sommers,
J. Phys. {\bf A
42} (2009) 012001
[arXiv:0810.1458].

\bibitem{APSoII}
G. Akemann, M.J. Phillips and H.-J. Sommers,
J. Phys. {\bf A
43} (2010) 085211
[arXiv:0911.1276].

\bibitem{Eynard}
B. Eynard,
J. Phys. {\bf A34} (2001) 7591
[arXiv:cond-mat/0012046].

\bibitem{GP} S. Ghosh and A. Pandey, Phys. Rev. {\bf E65} (2002) 046221. 


\bibitem{BHJ}
B.A. Khoruzhenko and H.-J. Sommers,
``Non-Hermitian Random Matrix Ensembles'',
invited chapter for the Oxford Handbook of Random Matrix Theory,
arXiv:0911.5645.


\bibitem{AP04}
G. Akemann and A. Pottier,
{J. Phys.}, {\bf A37} (2004) L453, 
[arXiv:math-ph/0404068].

\bibitem{Lehmann91} N. Lehmann and H.-J. Sommers,
  Phys. Rev. Lett. {\bf 67} (1991) 941. 

\bibitem{Edelman97} A. Edelman, J. Multivariate Anal. {\bf 60} (1997) 203.

\bibitem{Sinclair} C.D. Sinclair,
Int. Math. Res. Not. {\bf 2007} rnm015-15 (2007)
[arXiv:math-ph/0605006].

\bibitem{A07mu} J.J.M. Verbaarschot,
Les Houches Summer School, France, 6-25 June 2004,
arXiv:hep-th/0502029;\newline
G. Akemann,
Int. J. Mod. Phys. {\bf A22} (2007) 1077 [arXiv:hep-th/0701175].

\bibitem{Tilo} T.~Kanazawa, T.~Wettig and N.~Yamamoto,
Phys. Rev. \textbf{D81} (2010) 081701(R) 
[arXiv:0912.4999]. 

\bibitem{KGII} M. Kieburg and T. Guhr,
J. Phys. {\bf A
43} (2010) 135204
[arXiv:0912.0658].


\bibitem{AV03}
G. Akemann and G. Vernizzi,
  Nucl.\ Phys. {\bf B660} (2003) 532 [arXiv:hep-th/0212051].


\bibitem{Bergere}
  M.~C.~Berg\`ere,
  ``Biorthogonal Polynomials for Potentials of two Variables and External
  Sources at the Denominator'',
  arXiv:hep-th/0404126.


\bibitem{AB}
G. Akemann and F. Basile,
Nucl. Phys. {\bf B766} (2007) 150
[arXiv:math-ph/0606060].


\bibitem{FKS98} Y.V. Fyodorov, B.A. Khoruzhenko and H.-J. Sommers,
Ann. Inst. Henri Poincar\'e
{\bf 68} (1998) 449 [arXiv:chao-dyn/9802025].


\bibitem{A01} G. Akemann,
Phys. Rev. {\bf D64} (2001) 114021
[arXiv:hep-th/0106053].


\bibitem{Sommers2007} H.-J. Sommers,   J. Phys. {\bf A40} (2007) F671
[arXiv:0706.1671].

\bibitem{Sommers2008}
H.-J. Sommers and W. Wieczorek,  J. Phys.  {\bf A41} (2008) 405003
[arXiv:0806.2756].

\bibitem{KGI} M. Kieburg and T. Guhr,
J. Phys. {\bf A
43} (2010) 075201
[arXiv:0912.0654].



\bibitem{ForresterMays}
P.J. Forrester and A. Mays, 
J. Stat. Phys. {\bf 134} (2009) 443 
[arXiv:0809.5116].


\end{thebibliography}
\end{document}